\documentclass[aps,prd,preprintnumbers,showpacs,twocolumn,groupedaddress,nofootinbib]{revtex4}
\usepackage{graphicx}
\usepackage{latexsym}
\def\beq{\begin{equation}}
\def\eeq{\end{equation}}
\def\bey{\begin{eqnarray}}
\def\eey{\end{eqnarray}}

\def\sigv{\langle\sigma v\rangle}

\def\lsim{\mathrel{\raise.3ex\hbox{$<$\kern-.75em\lower1ex\hbox{$\sim$}}}}
\def\gsim{\mathrel{\raise.3ex\hbox{$>$\kern-.75em\lower1ex\hbox{$\sim$}}}}

\begin{document}

\title{On The Origin Of The Gamma Rays From The Galactic Center}  
\author{Dan Hooper$^{1,2}$ and Tim Linden$^{1,3}$}
\affiliation{$^1$Center for Particle Astrophysics, Fermi National Accelerator Laboratory, Batavia, IL 60510, USA}
\affiliation{$^2$Department of Astronomy and Astrophysics, University of Chicago, Chicago, IL 60637, USA}
\affiliation{$^3$Department of Physics, University of California, Santa Cruz, CA 95064, USA}

\date{\today}

\begin{abstract}

The region surrounding the center of the Milky Way is both astrophysically rich and complex, and is predicted to contain very high densities of dark matter. Utilizing three years of data from the Fermi Gamma Ray Space Telescope (and the recently available Pass 7 ultraclean event class), we study the morphology and spectrum of the gamma ray emission from this region and find evidence of a spatially extended component which peaks at energies between 300 MeV and 10 GeV. We compare our results to those reported by other groups and find good agreement. The extended emission could potentially originate from either the annihilations of dark matter particles in the inner galaxy, or from the collisions of high energy protons that are accelerated by the Milky Way's supermassive black hole with gas. If interpreted as dark matter annihilation products, the emission spectrum favors dark matter particles with a mass in the range of 7-12 GeV (if annihilating dominantly to leptons) or 25-45 GeV (if annihilating dominantly to hadronic final states). The intensity of the emission corresponds to a dark matter annihilation cross section consistent with that required to generate the observed cosmological abundance in the early universe ($\sigma v\sim 3 \times 10^{-26}$ cm$^3$/s). We also present conservative limits on the dark matter annihilation cross section which are at least as stringent as those derived from other observations.

\end{abstract}

\pacs{95.85.Pw,95.55.Ka,95.35.+d; FERMILAB-PUB-11-505-A}
\maketitle

\section{Introduction}

Since its launch in June of 2008, the Fermi Gamma Ray Space Telescope (FGST) has been producing the most detailed and highest resolution observations to date of the gamma ray sky between 50 MeV and 100 GeV. Among the objectives of this experiment are to increase our understanding of how astrophysical objects such as black holes and pulsars accelerate cosmic rays, and to help identify the substance or substances that compose the dark matter of our universe. For each of these areas of inquiry, the region surrounding the center of the Milky Way represents a particularly interesting and promising target of study. On the one hand, the Galactic Center is an extraordinarily rich and complex region, containing our galaxy's supermassive black hole, as well as supernova remnants, massive X-ray binary systems, massive O and B type stars, and two young and massive star clusters (Arches and Quintuplet)~\cite{astrogc,astrogc2,astrogc3,astrogc4}. On the other hand, the Galactic Center is predicted to contain very high densities of dark matter, which in many models leads to a very high rate of dark matter annihilation, and a correspondingly high luminosity of gamma rays. No other astrophysical source or region is expected to be as bright in dark matter annihilation products as the Galactic Center.

In this article, we follow previous work~\cite{HG2} and perform a detailed study of the spectral and morphological features of the gamma rays from the Galactic Center region, with the intention of identifying or constraining the origins of these particles. In particular, we produce gamma ray maps which reveal the presence of both a bright, approximately point-like, gamma ray source at the Galactic Center, along with a more spatially extended emission component. The spectrum of this extended source peaks strongly between several hundred MeV and $\sim$10 GeV. We find good agreement between our results and those reported by other groups~\cite{HG2,Boyarsky:2010dr,aharonian}.

In discussing the possible origins of this extended emission, we find again that the observed spectrum and morphology are consistent with that predicted from annihilating dark matter particles with a mass of 7-12 GeV annihilating dominantly to leptons~\cite{HG2} or a mass of 25-45 GeV annihilating dominantly to hadronic final states~\cite{HG1}. In either case, the normalization of the gamma ray flux requires an annihilation cross section that is consistent, within astrophysical uncertainties, with the value predicted for a simple thermal relic ($\sigma v \sim 3\times 10^{-26}$ cm$^3$/s). We also discuss the possibility that the extended gamma ray emission is produced through the collisions of energetic protons which are accelerated by the supermassive black hole with gas~\cite{aharonian}. While we consider this to be the leading astrophysical explanation for the gamma ray emission observed by the FGST, it is somewhat difficult to assess this hypothesis given how little is known or can be reliably predicted about the spectrum or flux of protons accelerated by the central black hole, and how little is known about the history of this object (such as periods of flaring and relative inactivity) and the properties of the surrounding interstellar medium.
 
The remainder of this article is structured as follows. In Sec.~\ref{analysis}, we describe our analysis of the Fermi data and present gamma ray maps of the Inner Galaxy and the corresponding spectrum of this emission. In Sec.~\ref{char} we further describe the properties of this emission and compare our results to those found by other groups. In Sec.~\ref{origin} we discuss several possible origins of this emission, including energetic protons from the central supermassive black hole, dark matter annihilations, and a population of gamma ray pulsars. In Sec.~\ref{con}, we derive constraints on the dark matter annihilation cross section which are at least as stringent as those based on other observations, such as those of dwarf spheroidals, galaxy clusters, the cosmological diffuse background, and nearby subhalos. In Sec.~\ref{summary}, we discuss our results within the larger context of dark matter searches and summarize our conclusions.

\section{Analysis Procedure}
\label{analysis}

\begin{figure*}[t]
\centering
\includegraphics[angle=0.0,width=2.45in]{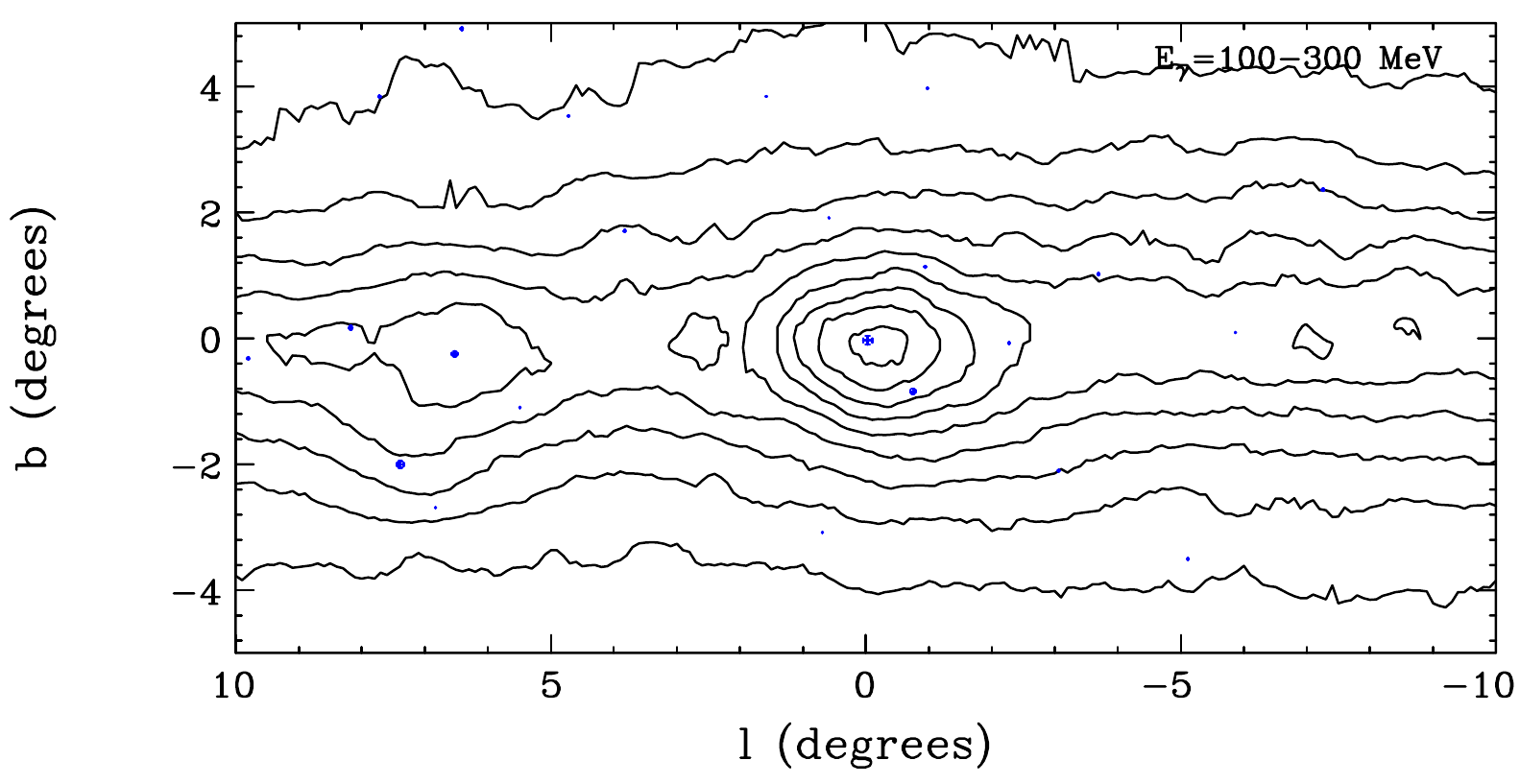}
\includegraphics[angle=0.0,width=2.2in]{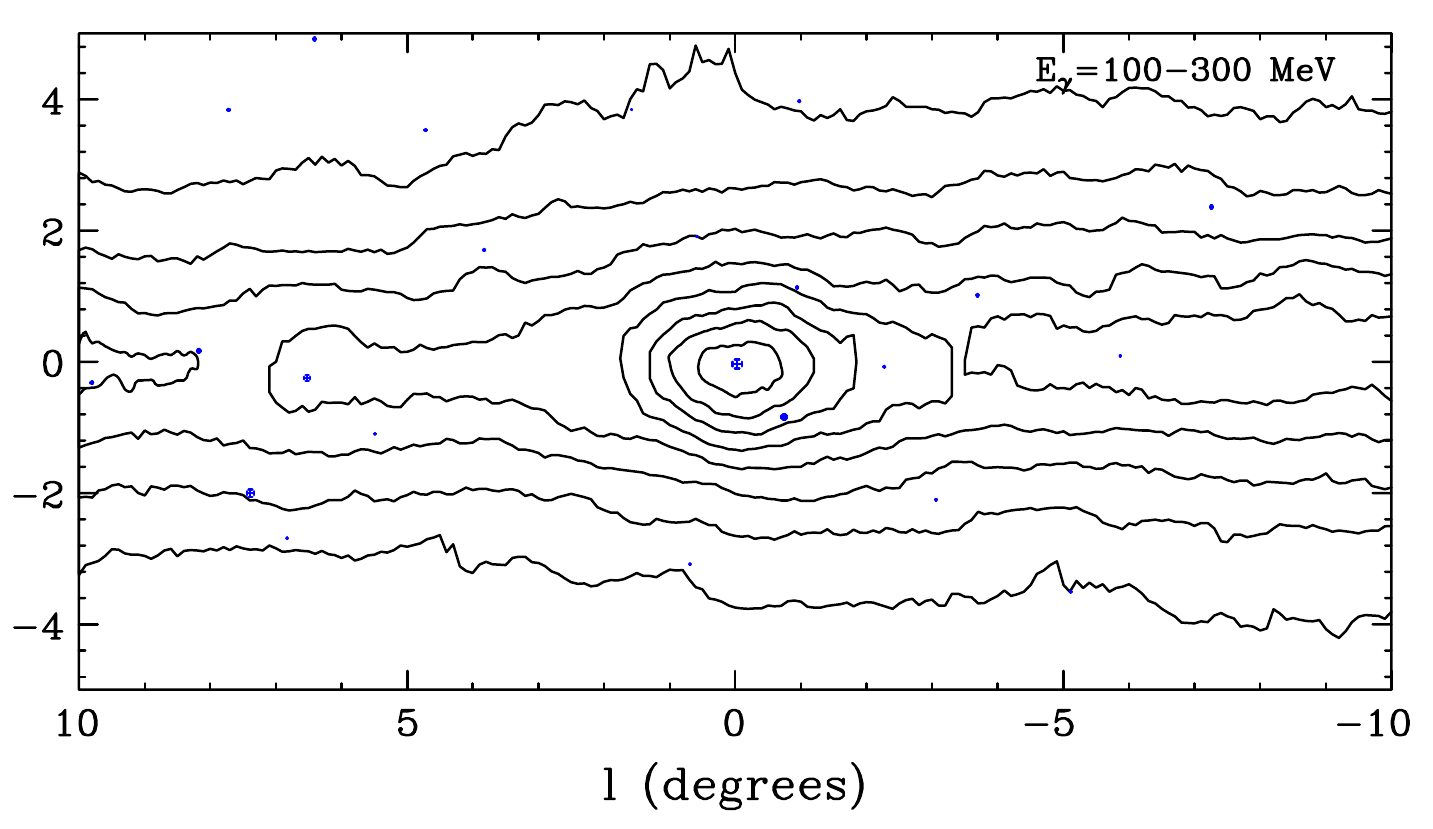}
\includegraphics[angle=0.0,width=2.2in]{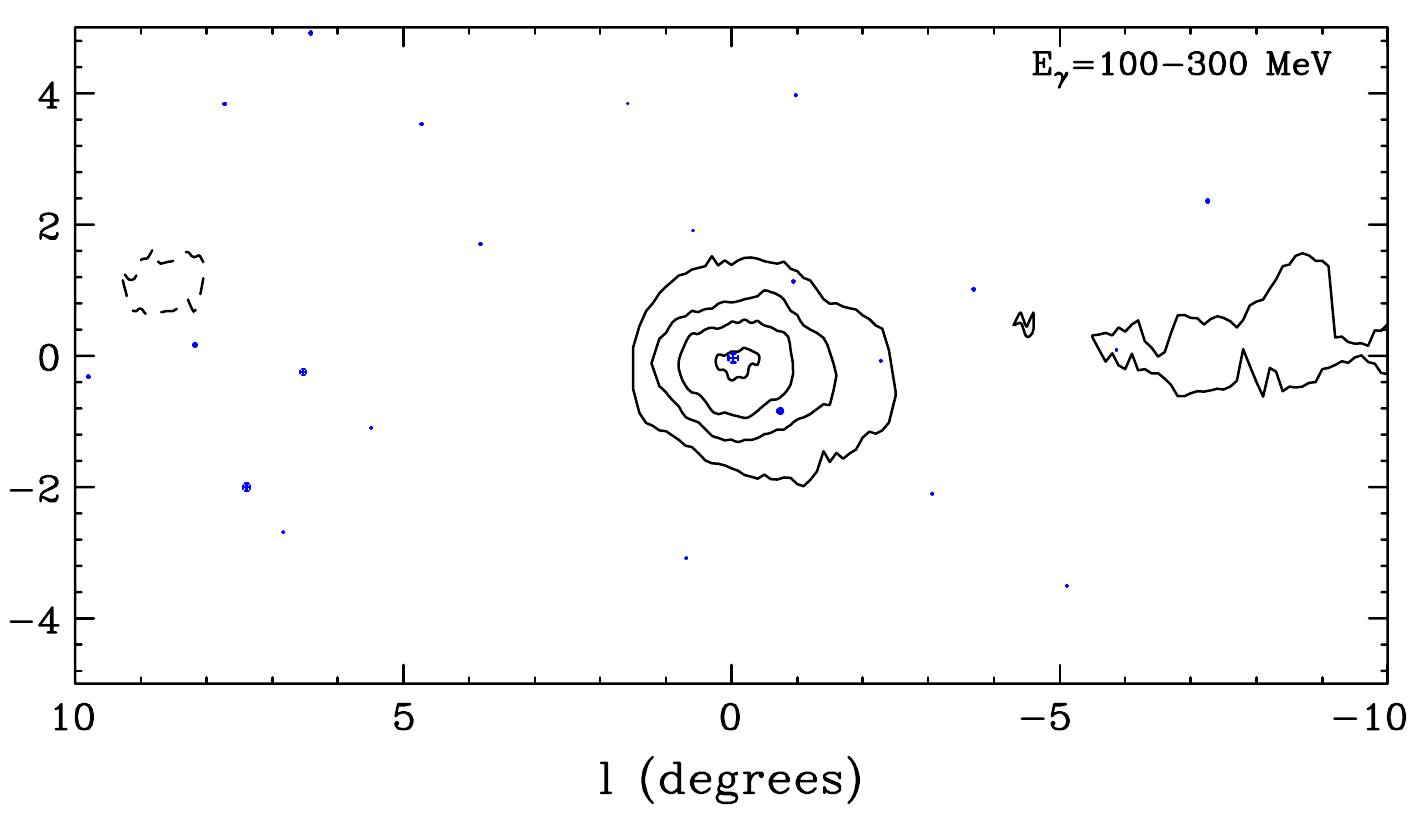}\\
\includegraphics[angle=0.0,width=2.45in]{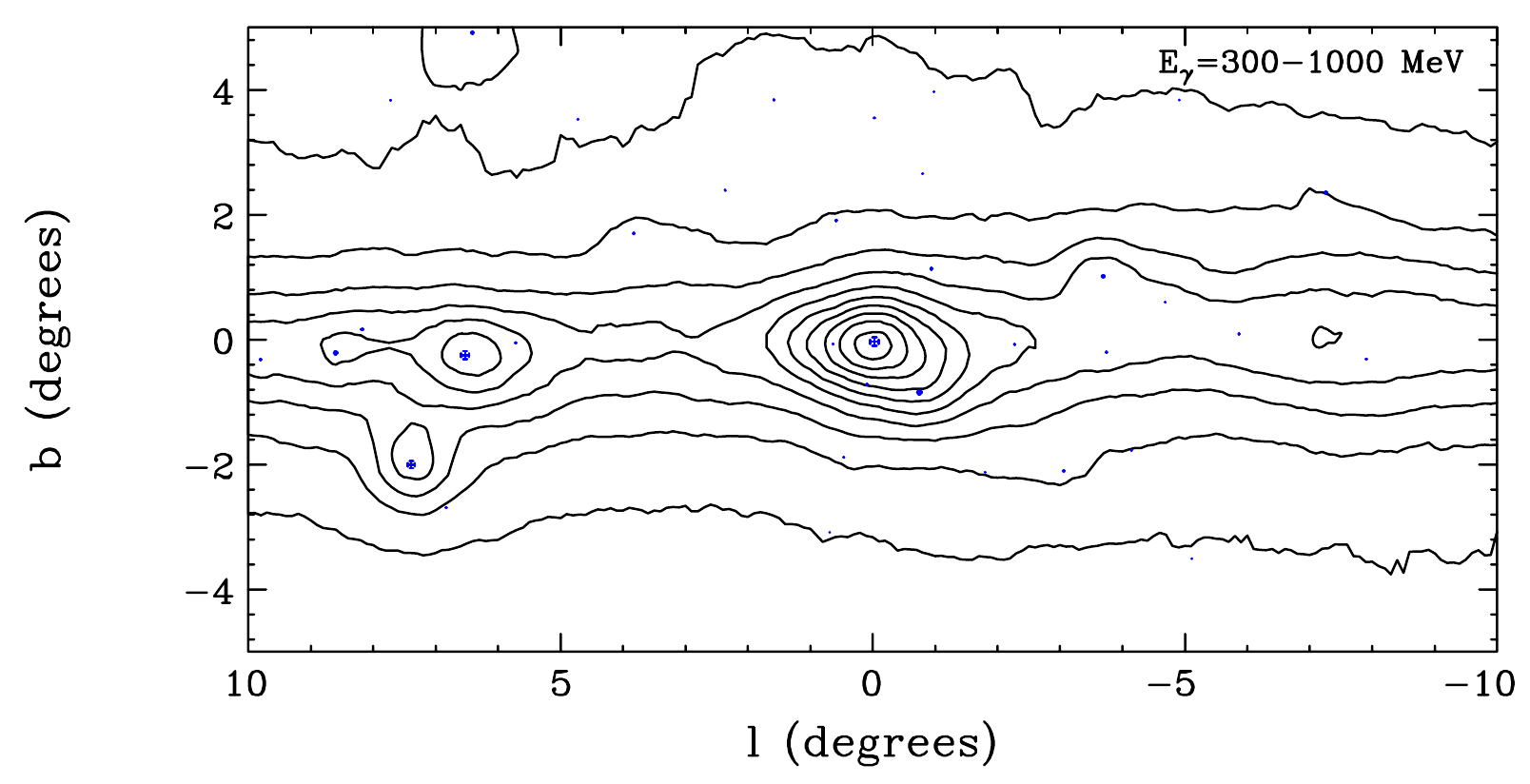}
\includegraphics[angle=0.0,width=2.2in]{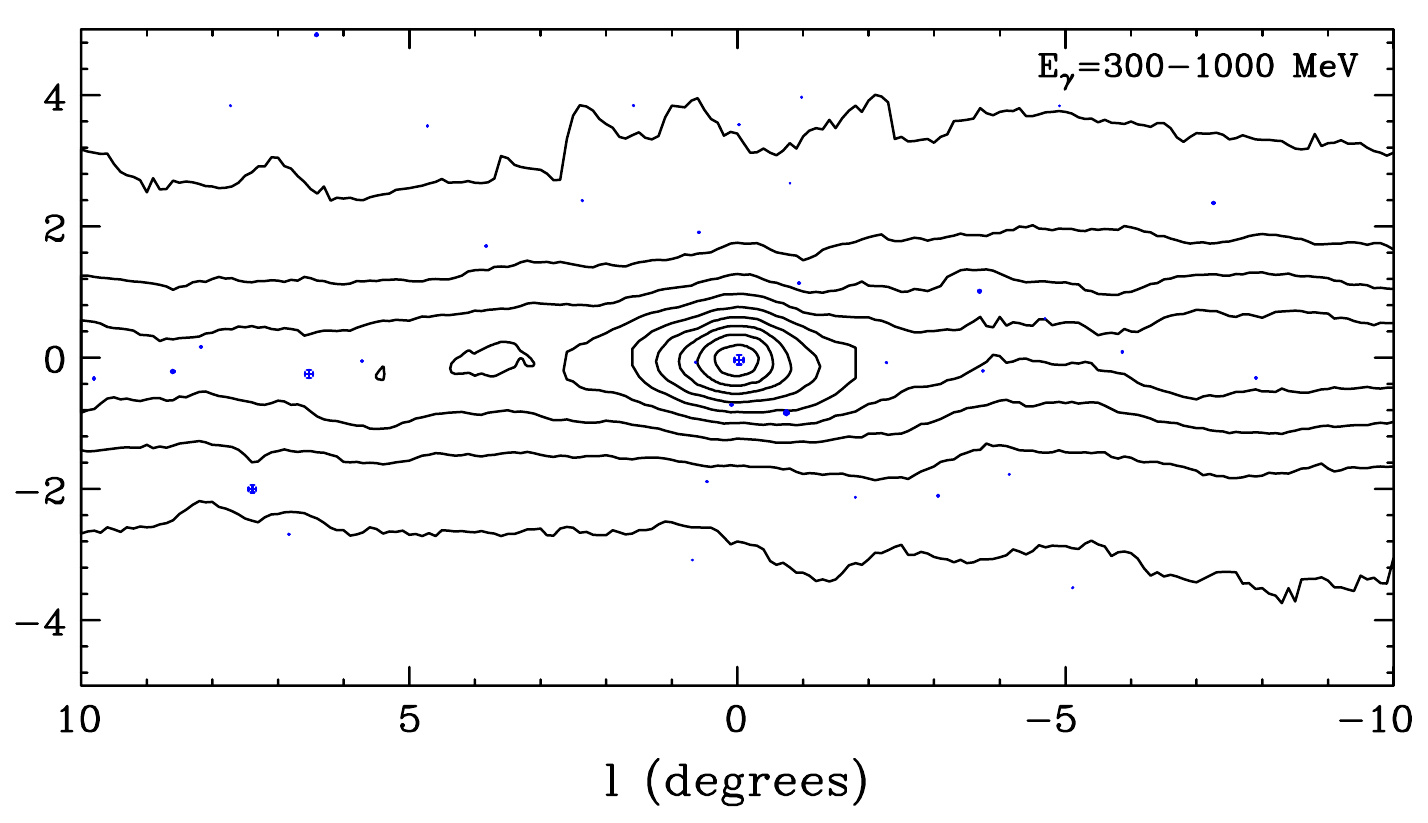}
\includegraphics[angle=0.0,width=2.2in]{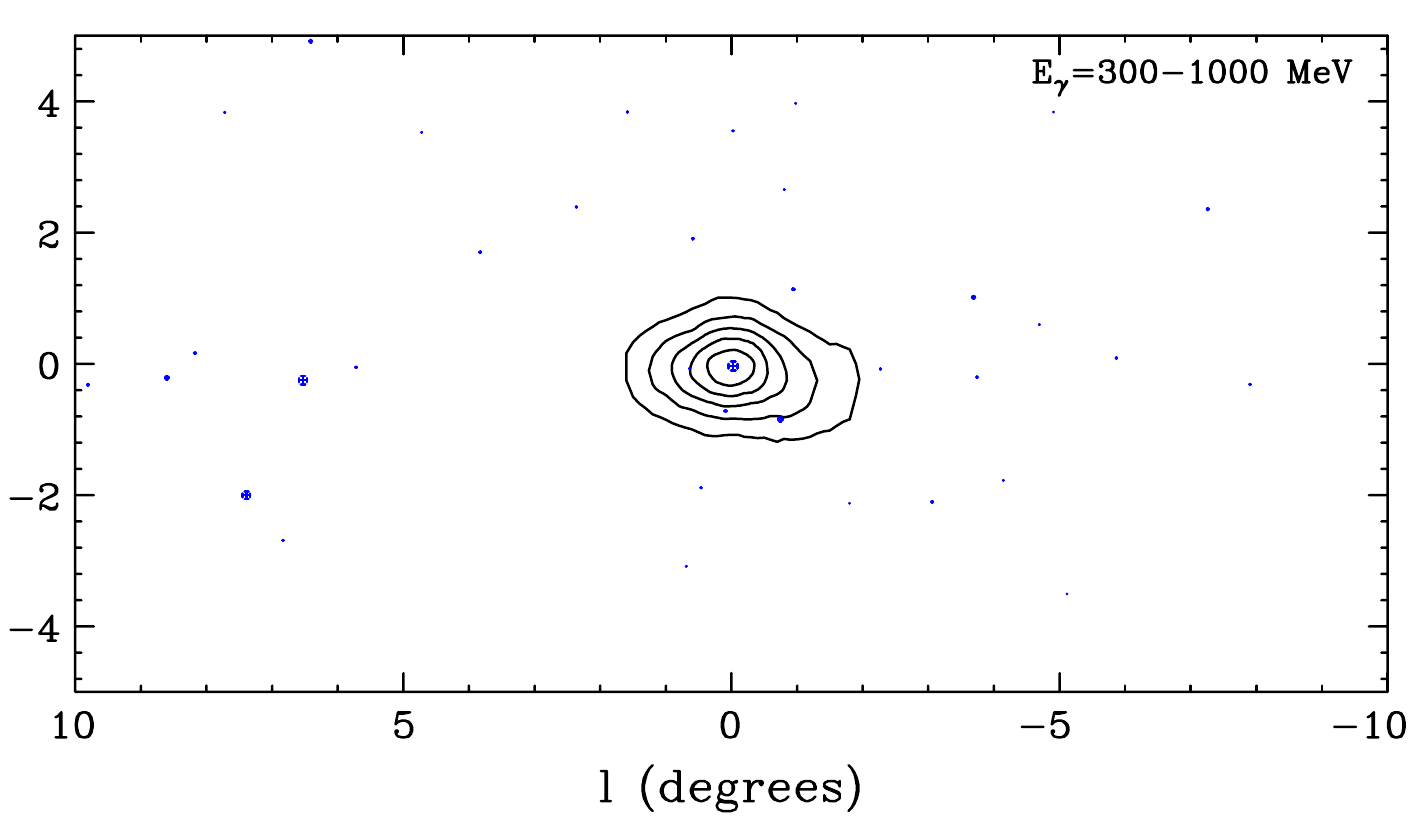}\\
\includegraphics[angle=0.0,width=2.45in]{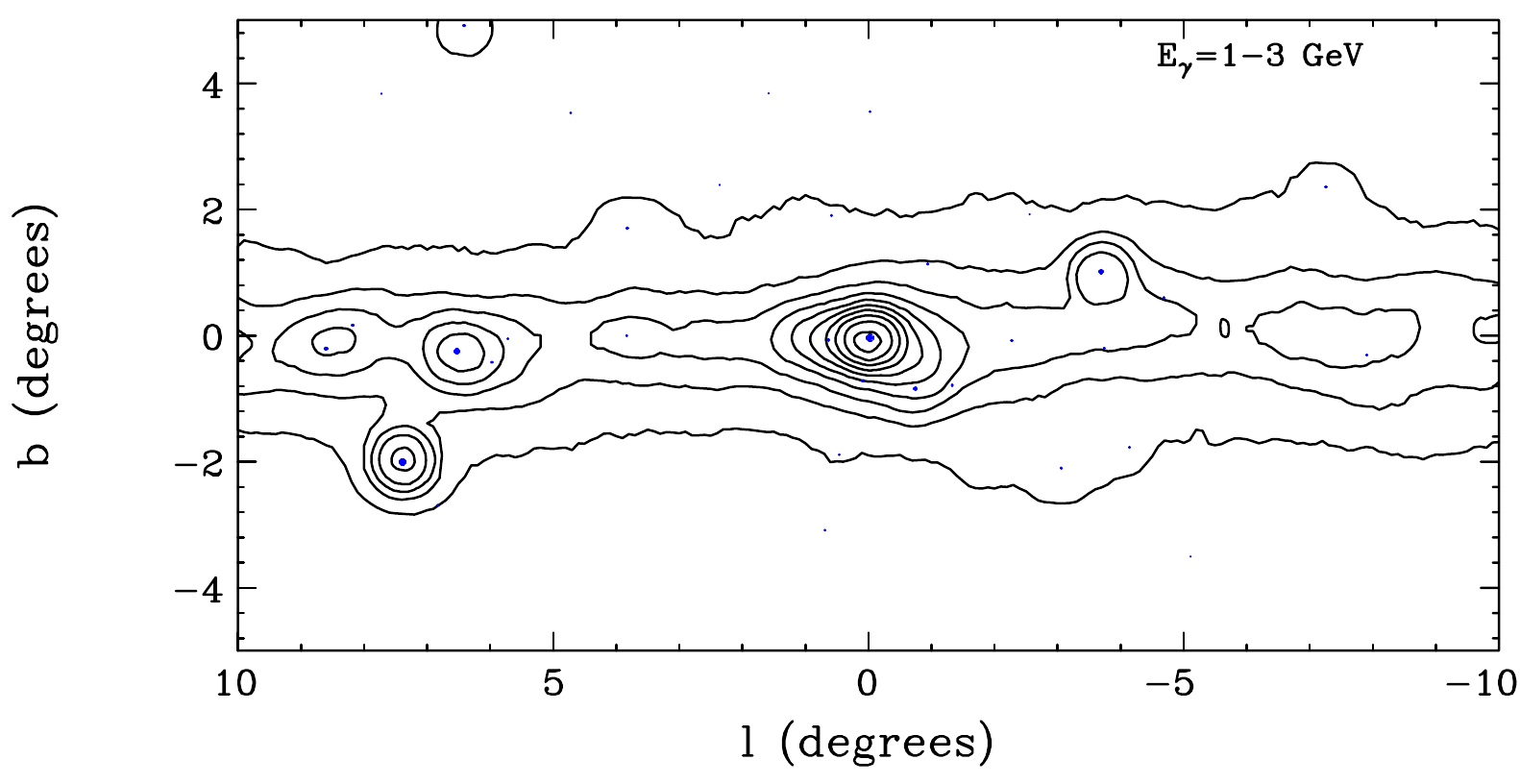}
\includegraphics[angle=0.0,width=2.2in]{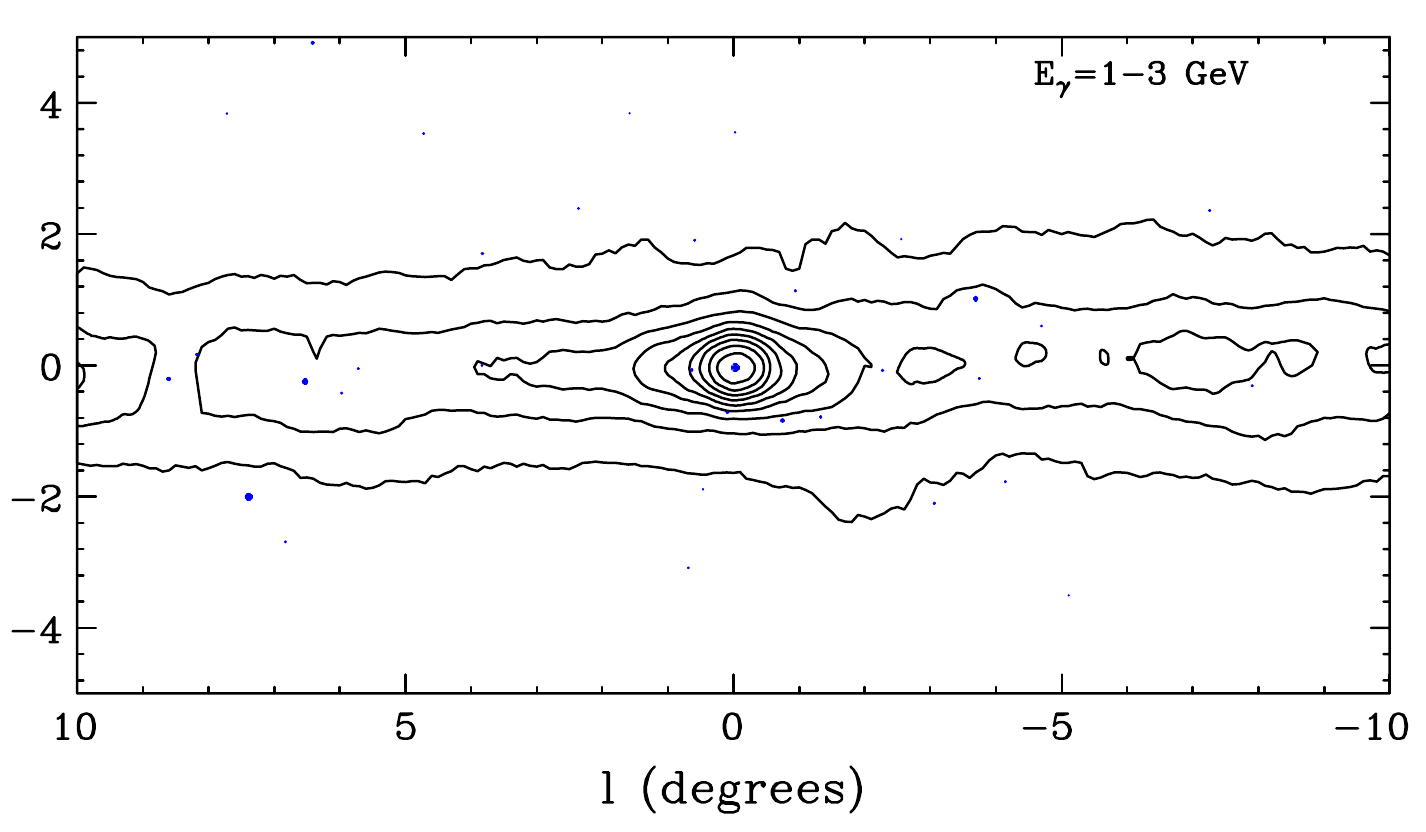}
\includegraphics[angle=0.0,width=2.2in]{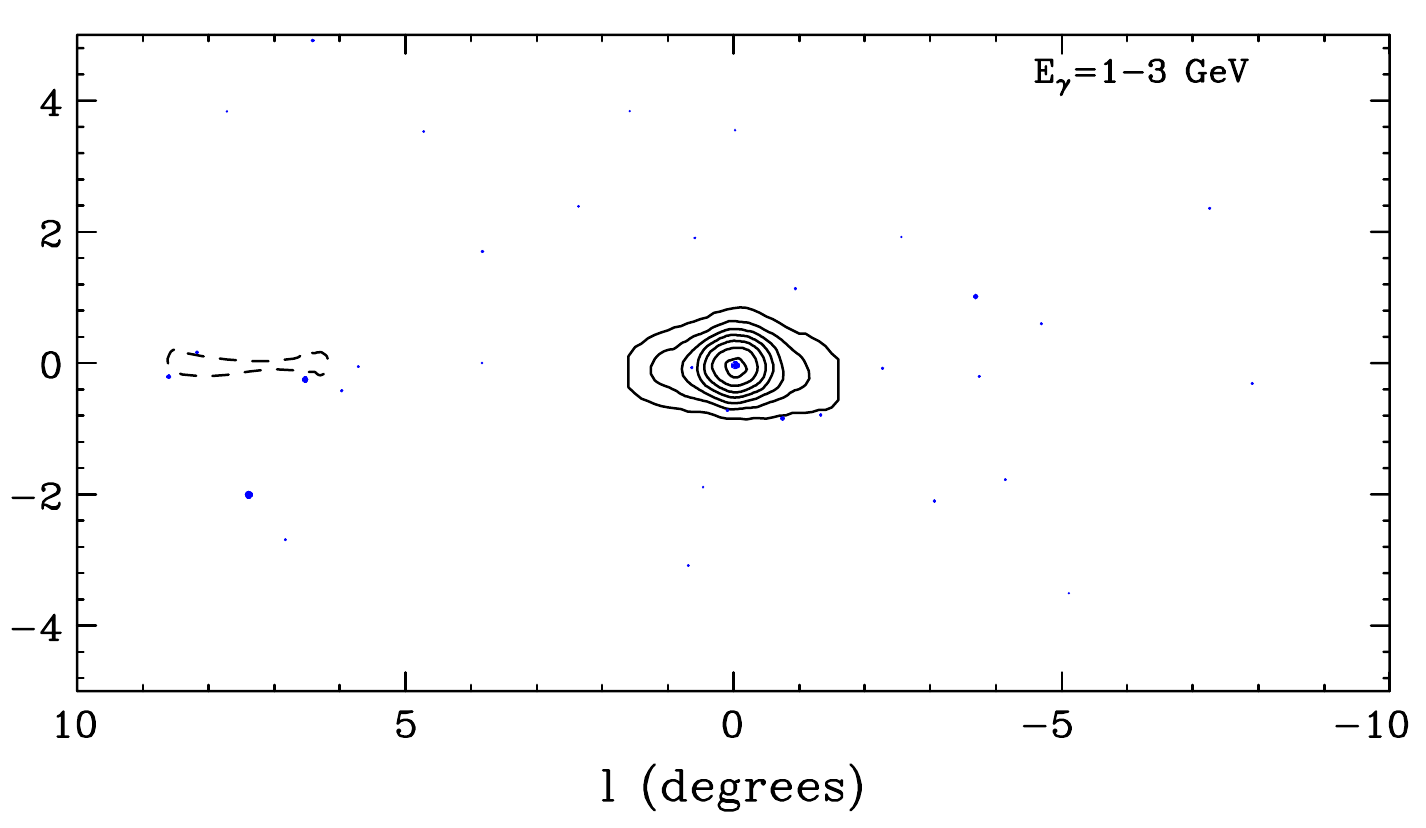}\\
\includegraphics[angle=0.0,width=2.45in]{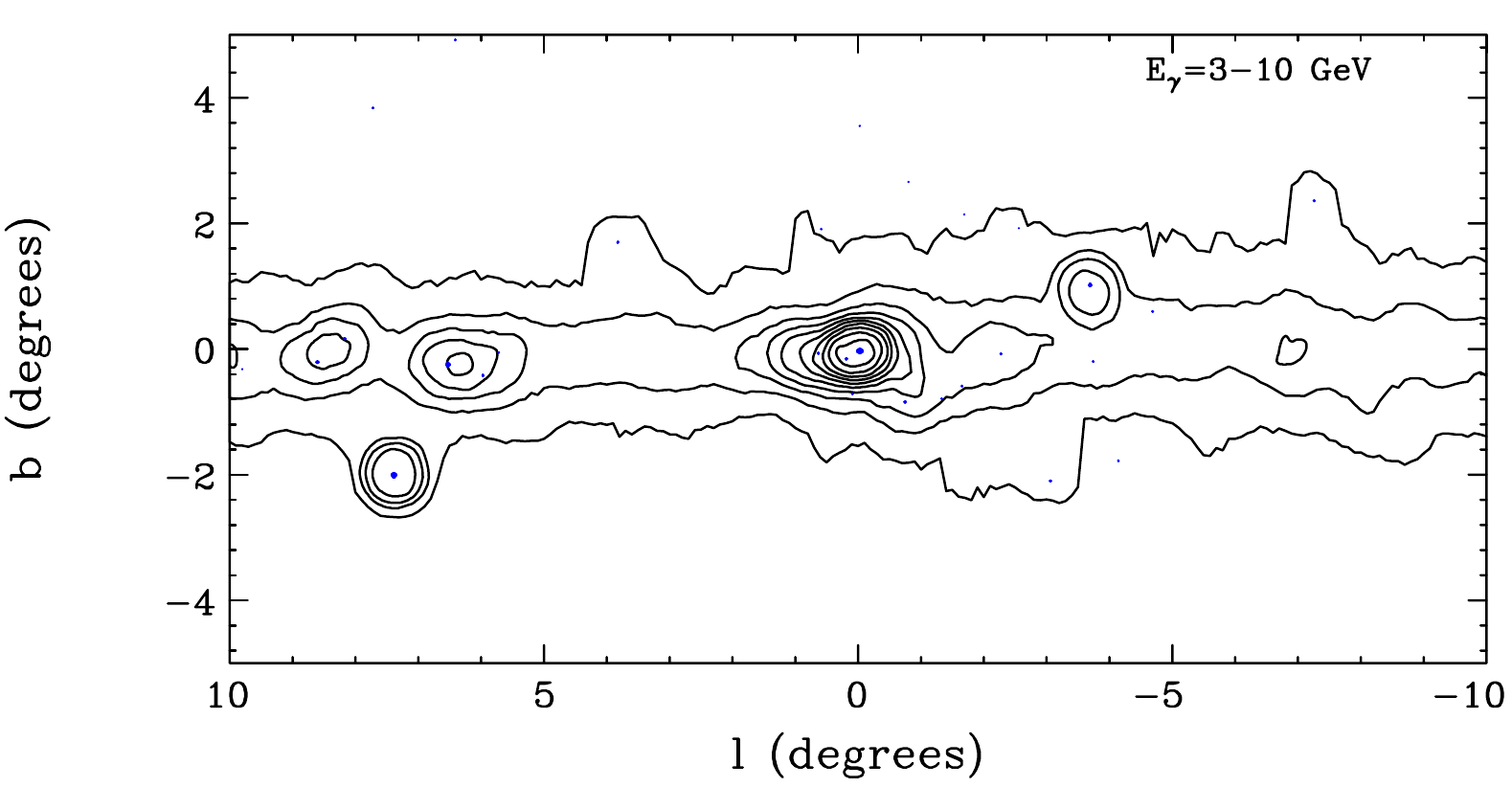}
\includegraphics[angle=0.0,width=2.2in]{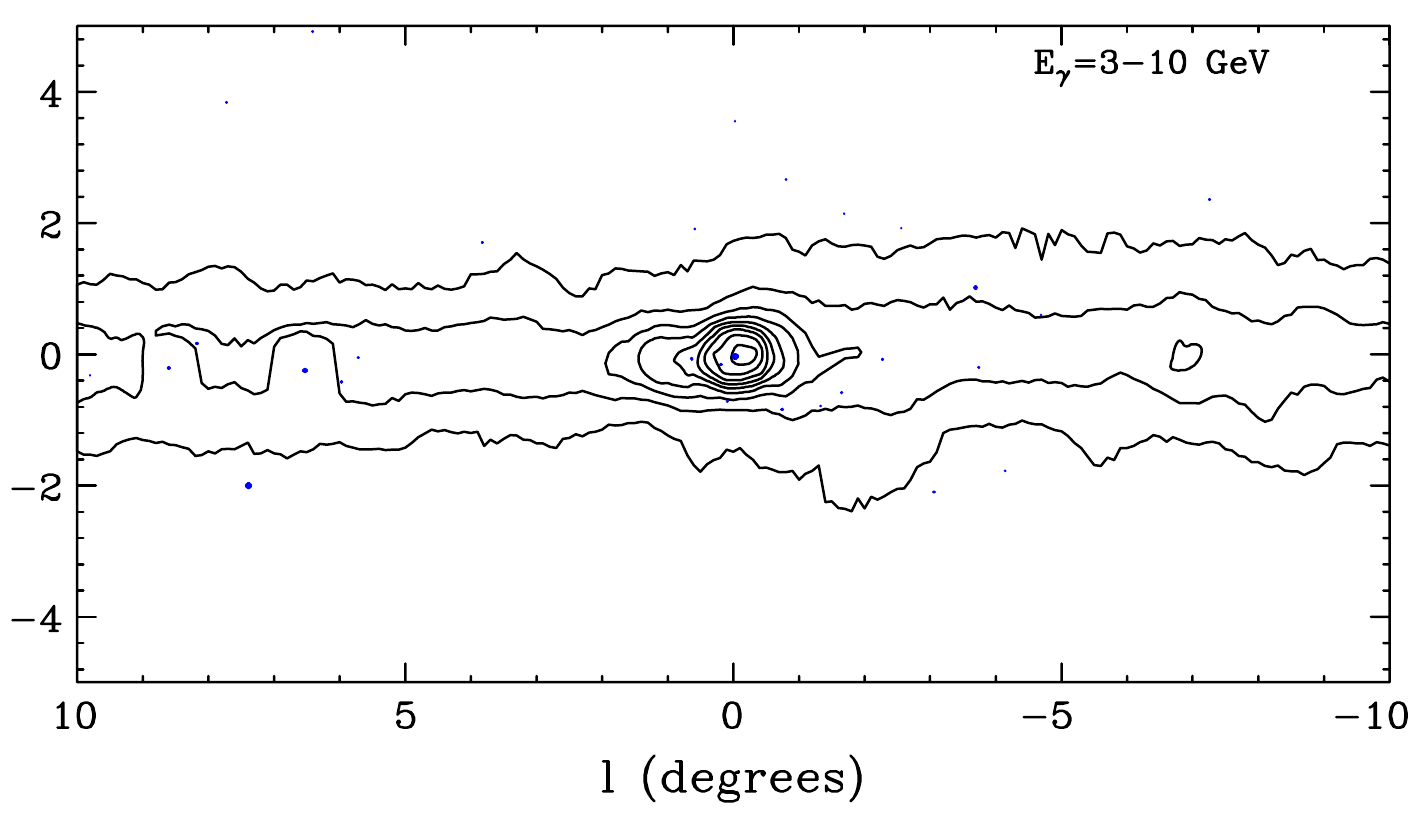}
\includegraphics[angle=0.0,width=2.2in]{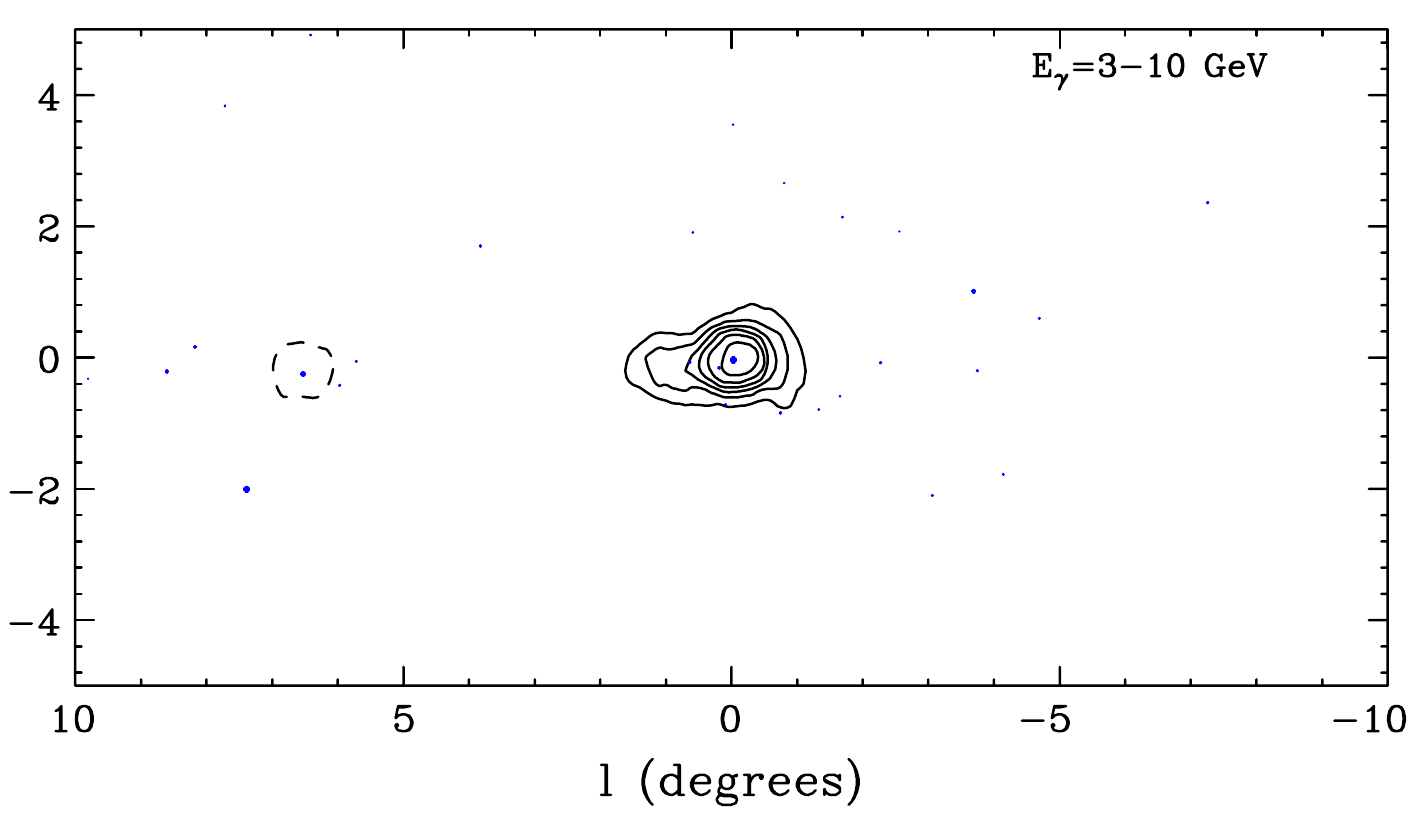}\\
\includegraphics[angle=0.0,width=2.45in]{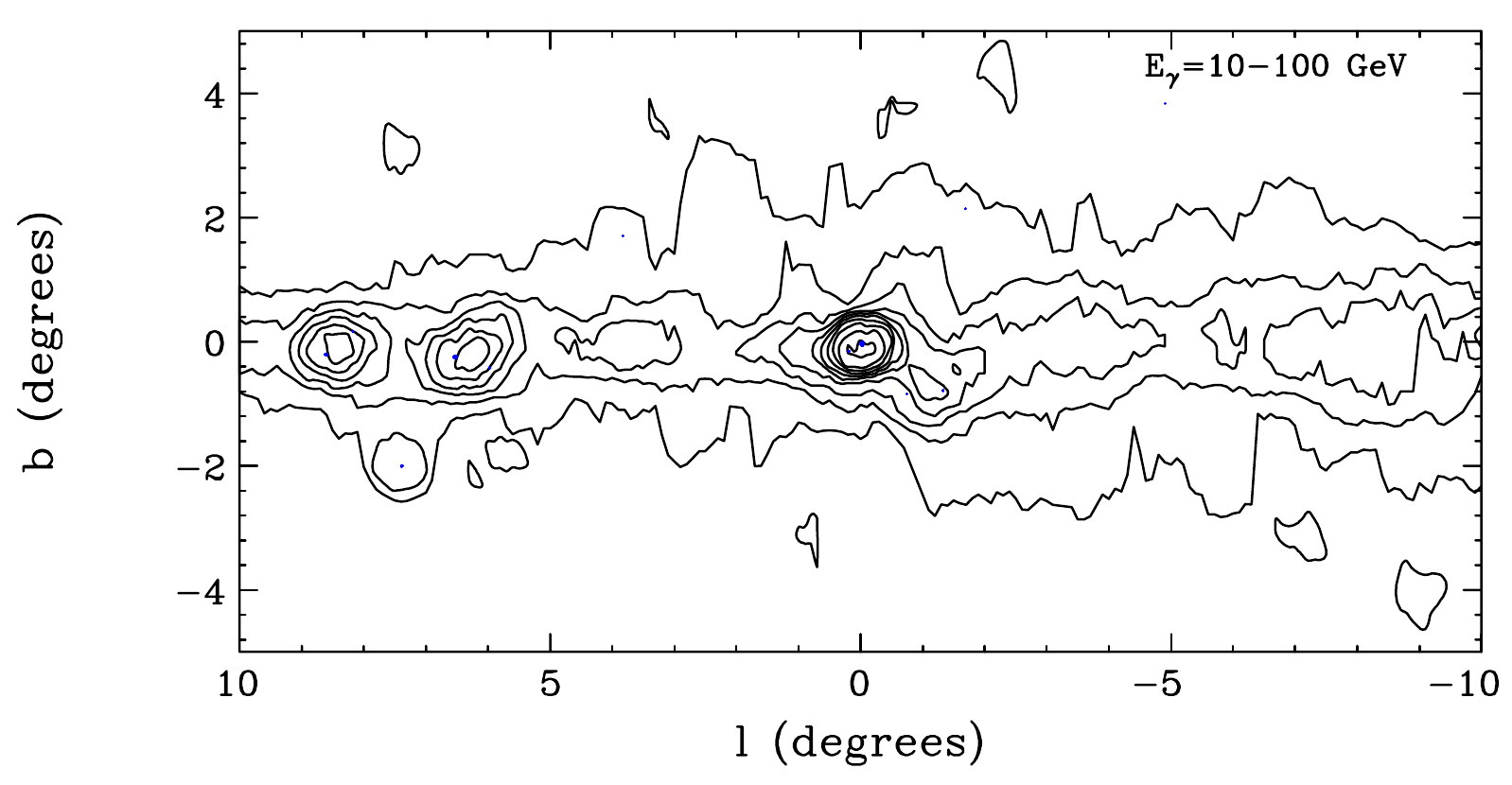}
\includegraphics[angle=0.0,width=2.2in]{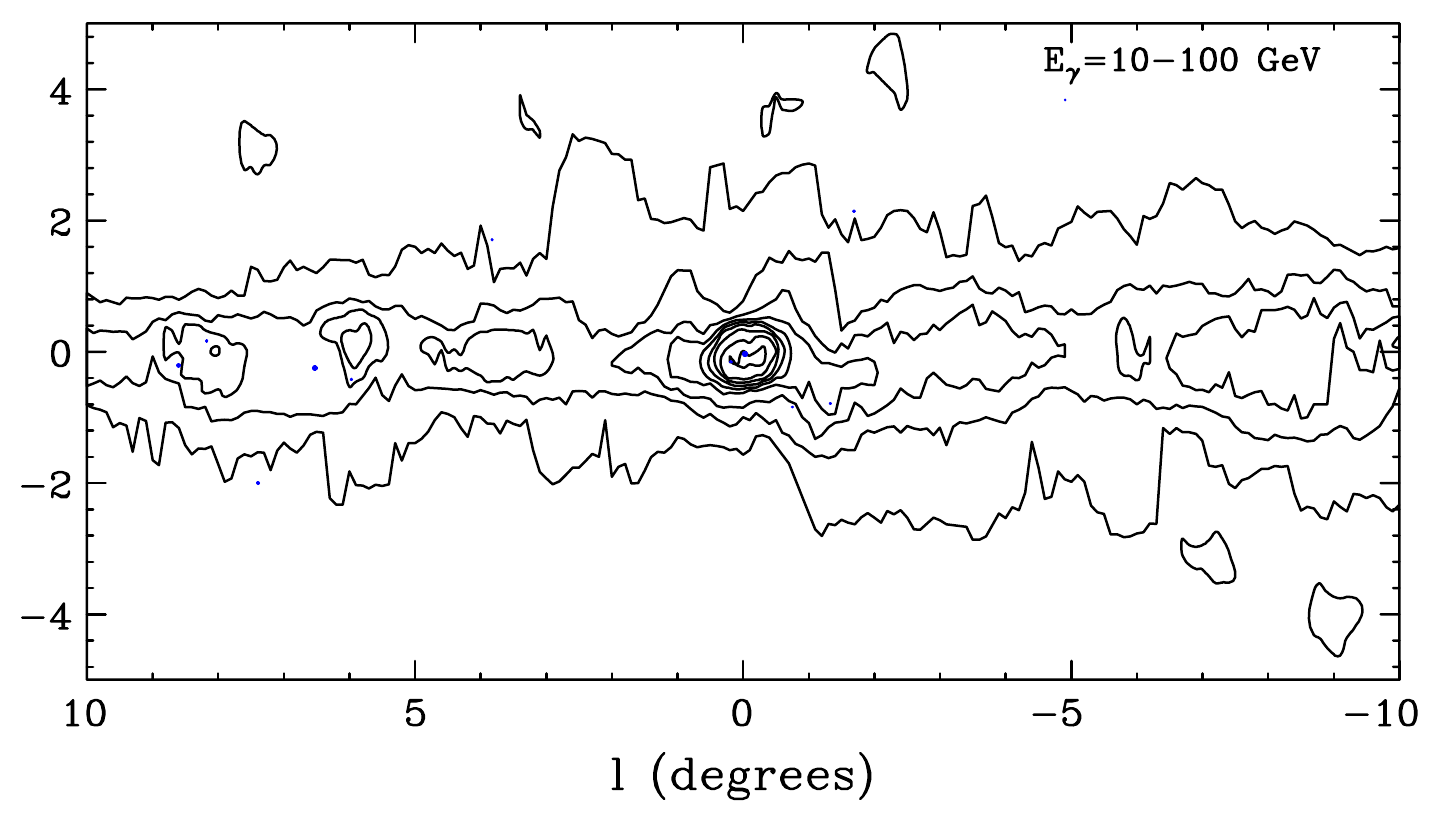}
\includegraphics[angle=0.0,width=2.2in]{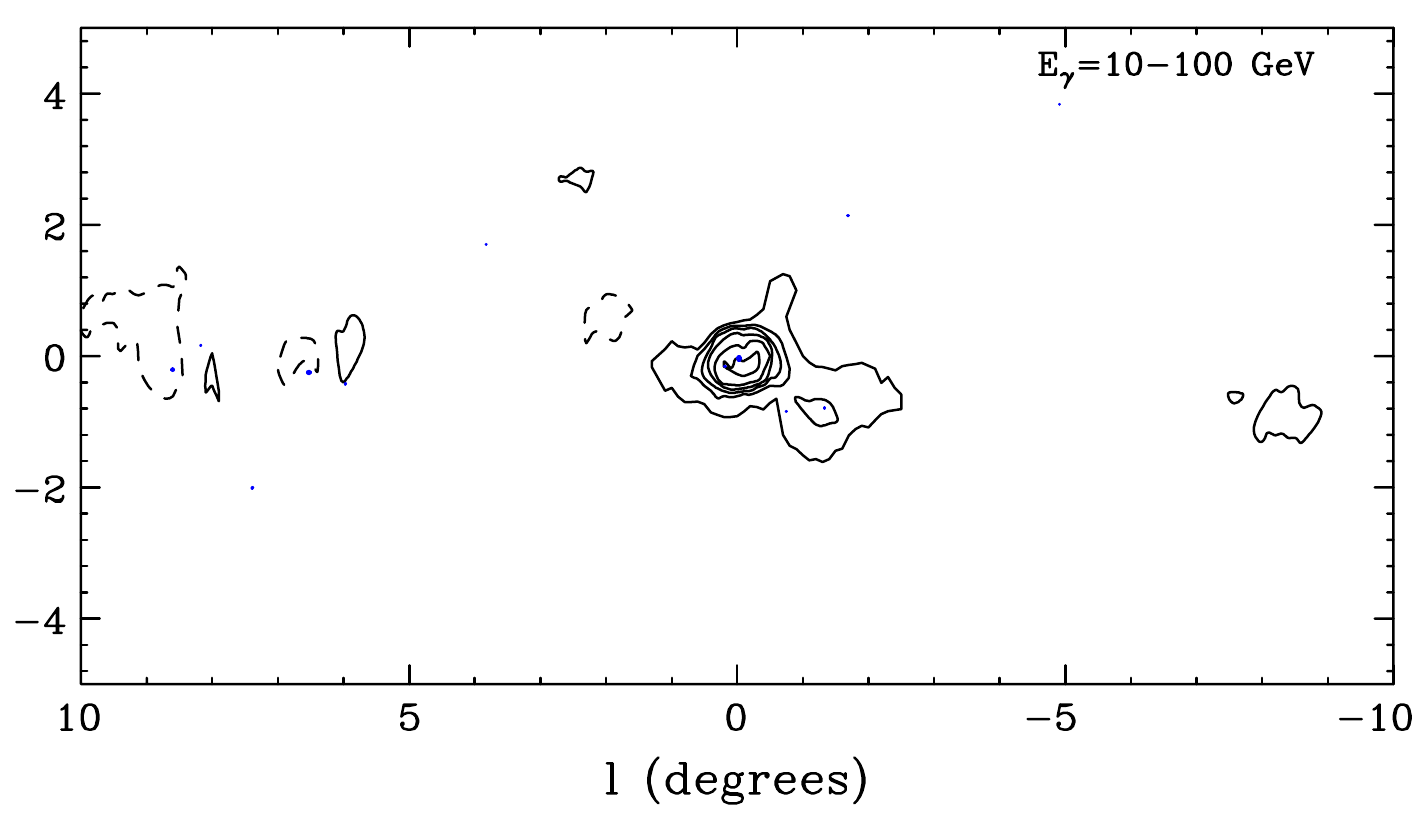}
\caption{Contour maps of the gamma ray flux from the region surrounding the Galactic Center, as observed by the Fermi Gamma Ray Space Telescope. The left frames show the raw maps, while the center and right frames show the maps after subtracting known sources (not including the central source), and known sources plus emission from cosmic ray interactions with gas in the Galactic Disk, respectively. All maps have been smoothed over a scale of 0.5 degrees. See text for more details.}
\label{maps}
\end{figure*}

We begin our analysis by generating contour maps of the region surrounding the Galactic Center which describe the distribution of gamma rays observed by the Fermi-LAT (Large Area Telescope) over the three years between August 4, 2008 and August 3, 2011. These maps were derived using only front-converting events (which have a superior point-spread function compared to back-converting events) from the Pass 7 ultraclean class. As recommended by the FGST collaboration, we include only events with zenith angles smaller than 100 degrees, and do not include events recorded while the Fermi satellite was transitioning through the South Atlantic Anomaly or while the instrument was not in survey mode. Each of the maps has been smoothed out at a scale of 0.5 degrees (the contour maps thus represent the flux observed within a 0.5 degree radius of a given direction in the sky). These raw maps are shown in the left frames of Fig.~\ref{maps}, for five different energy ranges between 100 MeV and 100 GeV.

In each map, ten contours are shown, distributed linearly between $2.64\times 10^{-8}$ and $2.64\times 10^{-7} {\rm cm}^{-2}\, {\rm s}^{-1}$ sq deg$^{-1}$ (100-300 MeV), $2.45\times 10^{-8}$ and $2.45\times 10^{-7} {\rm cm}^{-2}\, {\rm s}^{-1}$ sq deg$^{-1}$ (300-1000 MeV), $1.07\times 10^{-8}$ and $1.07\times 10^{-7} {\rm cm}^{-2}\, {\rm s}^{-1}$ sq deg$^{-1}$ (1-3 GeV), $2.66\times 10^{-9}$ and $2.66\times 10^{-8} {\rm cm}^{-2}\, {\rm s}^{-1}$ sq deg$^{-1}$ (3-10 GeV), and $3.77\times 10^{-10}$ and $3.77\times 10^{-9} {\rm cm}^{-2}\, {\rm s}^{-1}$ sq deg$^{-1}$ (10-100 GeV). Note that the $2.64 \times 10^{-8} {\rm cm}^{-2}\, {\rm s}^{-1}$ sq deg$^{-1}$ contour appears out of the field in the upper-left and upper-middle frames.


The blue points shown in the maps represent the locations of sources contained in the Fermi Second Source Catalog (2FGL)~\cite{catalog}, and the size of each point is proportional to the reported intensity of the source in the energy range shown. To account for these sources, we have generated a template map of their emission (assuming the central values for their intensity and locations as reported in the 2FGL), and taking into account the point-spread function of the Fermi-LAT (as determined by the Fermi Tool gtpsf). In the center frames of Fig.~\ref{maps}, we show the maps as they appear after subtracting this source template. Note that we have not removed the central bright source, as its emission is difficult to disentangle from dark matter annihilation products originating from the inner region of a cusped halo profile. We will return to this issue later in the article.

\begin{figure*}[t]
\centering
\vspace{-1.0cm}
\includegraphics[angle=0.0,width=5.5in]{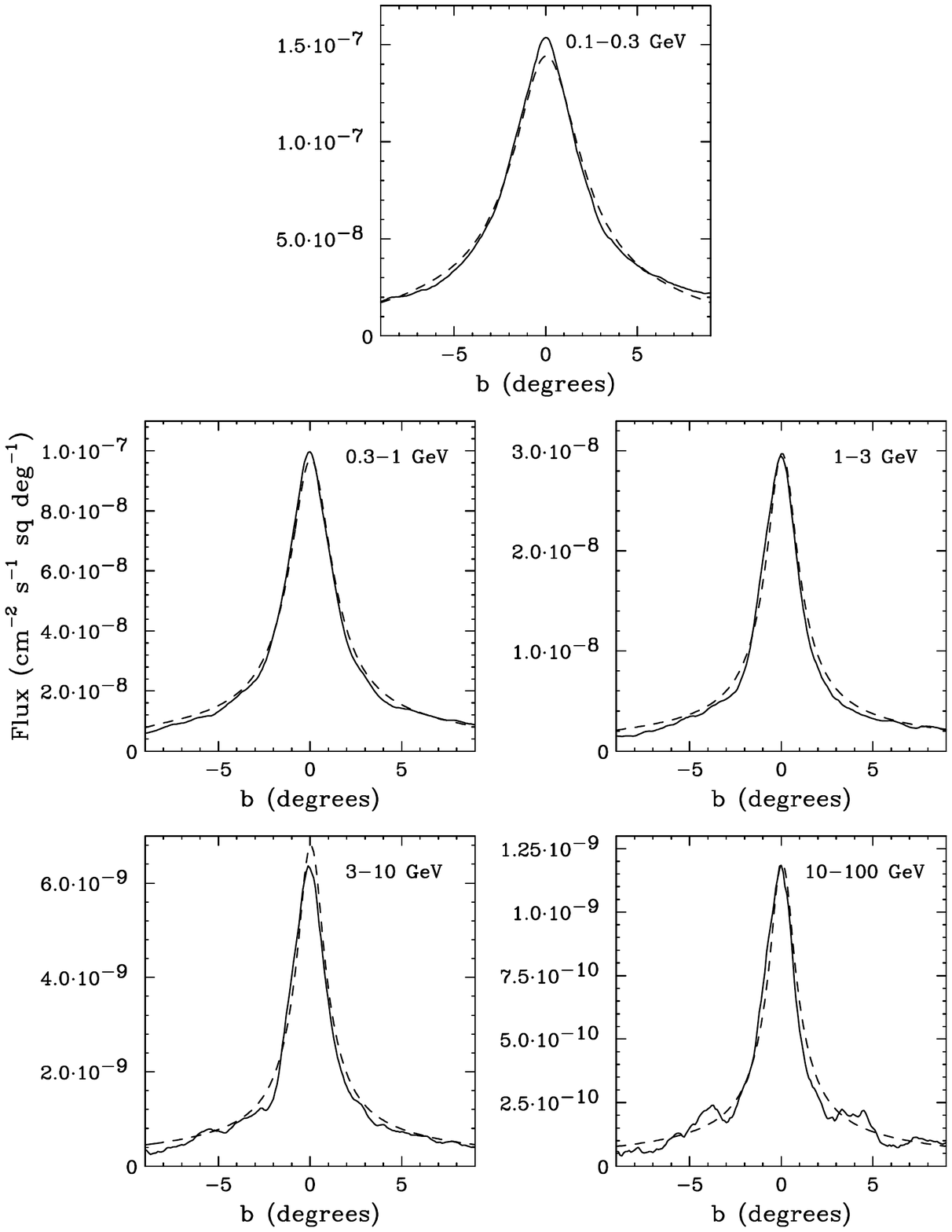}
\vspace{-2.3cm}
\caption{The observed gamma ray flux (after subtracting known sources) averaged over $7^{\circ}<|l|<10^{\circ}$ as a function of galactic latitude (solid), compared to that predicted from the line-of-sight gas density (dashes). See text for details.}
\label{disk}
\end{figure*}

After subtracting these known sources, the dominant remaining component is the diffuse emission associated with the disk of our galaxy. This emission is dominated by cosmic ray processes taking place throughout the disk of the Milky Way, which one must look through in order to observe the Galactic Center. By studying the morphology of this emission over the regions of $5^{\circ}<|l|<10^{\circ}$, we find only a modest degree of variation with galactic longitude. In Fig.~\ref{disk}, we show as solid lines the observed gamma ray flux as a function of galactic latitude, averaged over the range of $7^{\circ}<|l|<10^{\circ}$ (in order to avoid any contamination with emission from the inner most degrees, we do not here make use of the data within $7^{\circ}$). 

The gamma ray emission from the disk of our galaxy is dominated by the decays of neutral pions produced in cosmic ray interactions with gas, although inverse compton and bremsstrahlung components also contribute. To model the morphology of the pion decay component, we adopt the following distribution of gas~\cite{gas,gas2}:
\begin{eqnarray}
\rho_{\rm gas} &\propto& e^{-|z|/z_{\rm sc}(R)}, \,\,\,\,\,\,\,\,\,\,\,\,\,\,\,\,\,\,\,\,\,\,\,\,\,\,\, {R < 7\, {\rm kpc}}, \\
\rho_{\rm gas} &\propto& e^{-|z|/z_{\rm sc}(R)} \,  e^{-R/R_{\rm sc}},\,\,\,\,\, {R > 7\, {\rm kpc}},\nonumber
\end{eqnarray}
where $z$ and $R$ describe the location relative to the Galactic Center in cylindrical coordinates. We set $R_{\rm sc}=3.15$ kpc (as fit to the data shown in Fig.~4 of Ref.~\cite{gas}) and $z_{\rm sc}(R) = 0.1 + 0.00208 \times (R/{\rm kpc})^2$ kpc (as fit to Fig.~4 of Ref.~\cite{gas2}), in good agreement with observations of 21-cm surveys, which trace the density of neutral hydrogen. To estimate the flux of pion decay gamma rays, we integrate this distribution over the line-of-sight (and again smooth over a radius of 0.5 degrees). After accounting for the Fermi-LAT point-spread function, we find that this gas distribution leads to the morphology described by the dashed lines in Fig.~\ref{disk}. This is in good agreement with the observed morphology of the diffuse emission. We also note that the spectral shape implied by the relative fluxes in these five energy bins is consistent with that predicted for a combination of pion decay and inverse compton scattering processes, as previously found in Ref.~\cite{HG2}. By subtracting this disk template from the gamma ray maps, we are able to remove the overwhelming majority of the diffuse astrophysical background from our maps. We emphasize that in performing this subtraction, we are not extrapolating any physical features of the inner galaxy, but are merely extrapolating the line-of-sight gas densities along the disk from directions slightly away from the Galactic Center to those more aligned with the Galactice Center.

In the right frames of Fig.~\ref{maps}, we show the resulting maps after subtracting both the known sources template (again, not including the bright central source) and the line-of-sight gas template. In each energy range, the majority of the background has been accurately removed by this simple subtraction. While this subtraction procedure does not perfectly remove all likely astrophysical backgrounds, the residuals outside of the inner $\sim$2$^{\circ}$ are very modest, typically on order of 10\% or less of the residual flux in the innermost region of the Galaxy. We include the observed spatial variations of the residuals as a systematic error, which we propagate throughout this study.

The residuals in this innermost region include a roughly spherically symmetric component centered around the Galactic Center, along with a sub-dominant component that is somewhat extended along the disk. Due to its similar angular extent, we consider it likely that this component is associated with emission from proton-proton collisions taking place in the Galactic Ridge, as observed at higher energies by HESS~\cite{ridge}. The remaining spherically symmetric component could plausibly originate from dark matter annihilations, processes associated with the Milky Way's supermassive black hole, gamma ray pulsars, or a combination of these and other sources. We will return to these issues in Secs.~\ref{char} and \ref{origin}.

\begin{figure}[t]
\centering
\vspace{0.0cm}
\includegraphics[angle=0.0,width=3.5in]{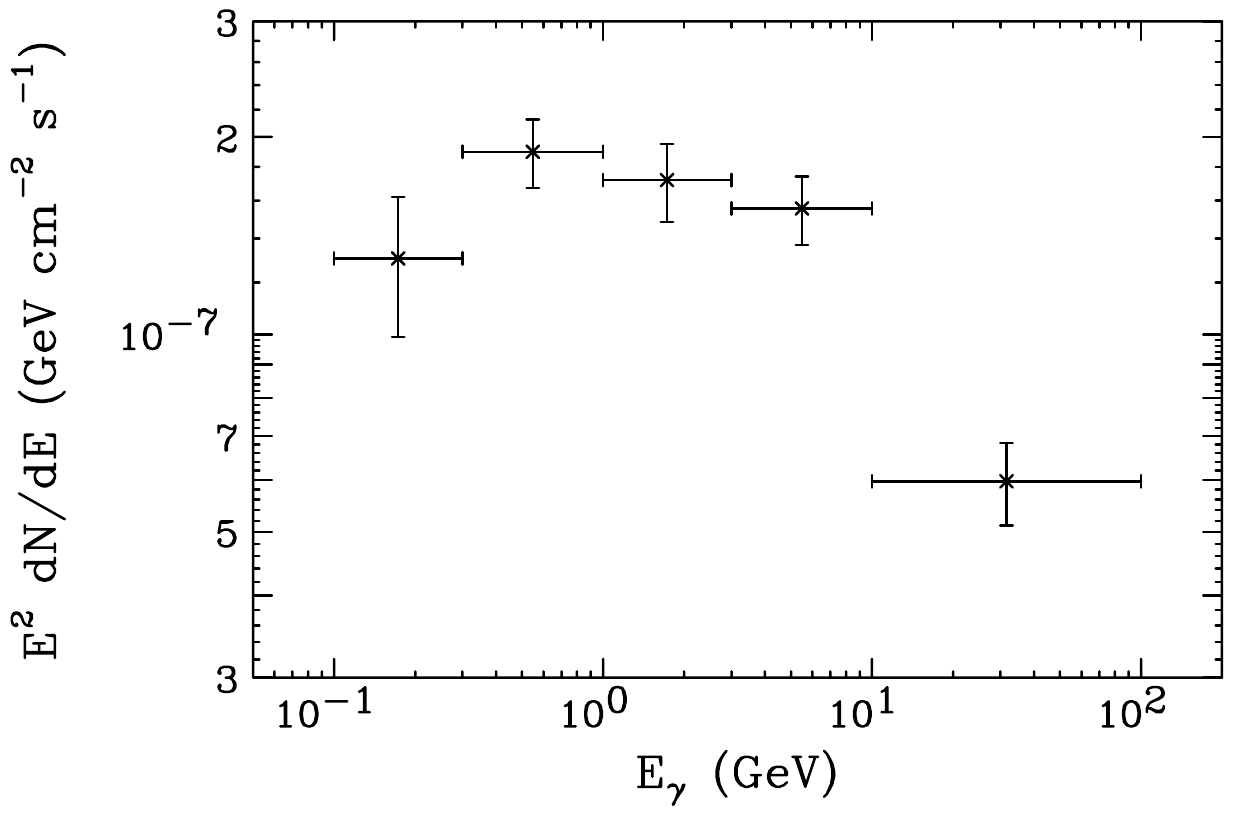}
\caption{The spectrum of the residual emission from the inner 5 degrees surrounding the Galactic Center, after subtracting the known sources and line-of-sight gas templates.}
\label{spec}
\end{figure}

In Fig.~\ref{spec}, we show the spectrum of the emission from the inner 5 degrees surrounding the Galactic Center, after removing the known sources and disk emission templates. The spectrum is clearly brightest between 300 MeV and 10 GeV, and drops by nearly an order of magnitude above $\sim$10 GeV. Note that the spectral shape of this residual is very similar to that (preliminarily) reported in conference presentations by the Fermi Collaboration~\cite{prelim}. In the following sections, we will explore the characteristics of this residual emission and discuss its possible origins.


\section{Properties of the Inner Emission}
\label{char}

\begin{figure}[t]
\centering
\vspace{0.0cm}
\includegraphics[angle=0.0,width=3.5in]{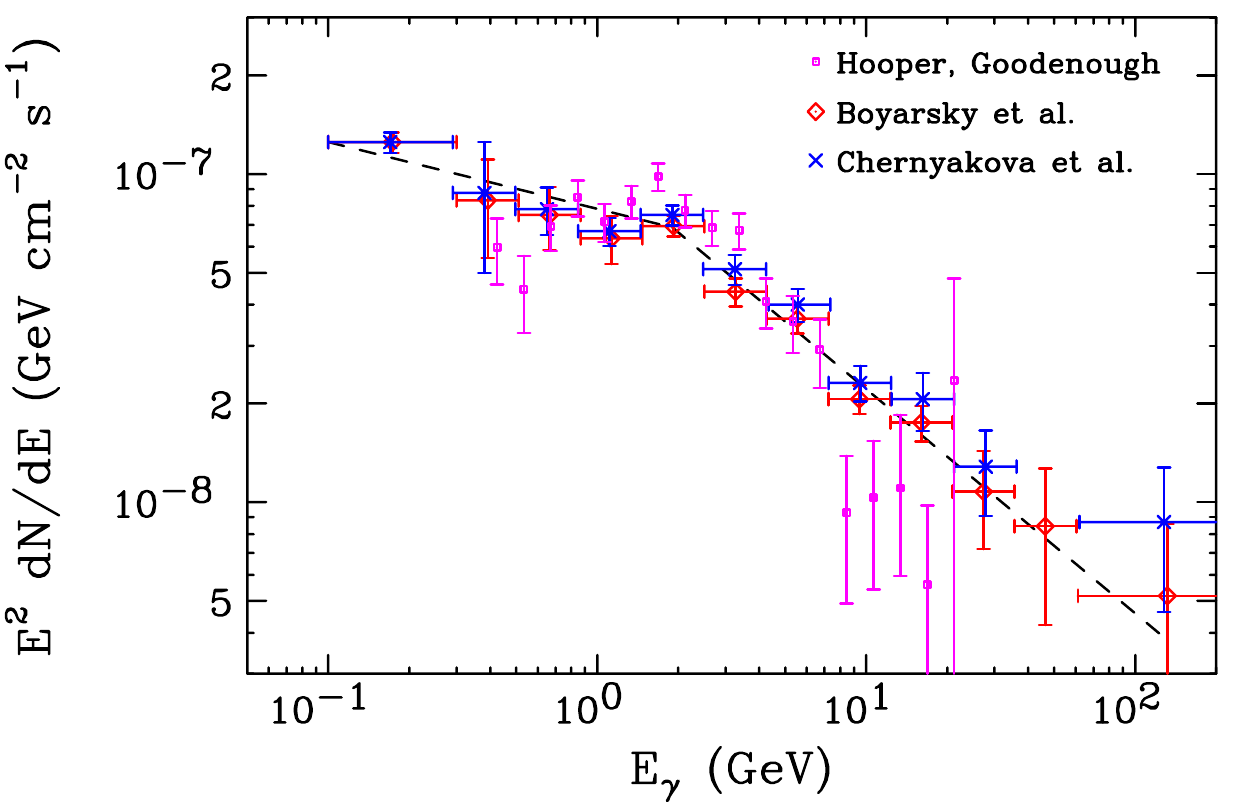}
\caption{Fits for the spectrum of the central emission, assuming a point-like source morphology, from the previous work of three different groups~\cite{HG2,Boyarsky:2010dr,aharonian}. Despite the different analysis approaches taken, these fits are all in reasonable agreement. The dashed line is the broken power-law fit to this spectrum as presented in Ref.~\cite{aharonian}.}
\label{pointspec}
\end{figure}

In an effort to constrain the origin (or origins) of the gamma rays from the inner region of our galaxy, we discuss in this section the spectral and morphological characteristics of the observed emission~\cite{method}. We begin by noting that the morphology of the observed residual is not consistent with that of a single point source. In particular, we find that above 300 MeV, less than half of the residual emission shown in the right frames of Fig.~\ref{maps} can be accounted for by a single, centrally-located point source. This conclusion is also supported by the independent analyses by Boyarsky {\it et al.}~\cite{Boyarsky:2010dr}, Chernyakova {\it et al.}~\cite{aharonian}, and Hooper and Goodenough~\cite{HG2}, which each found a spectrum of point-like emission from the Galactic Center which is considerably less intense than the total residual emission shown in Fig.~\ref{spec}. In Fig.~\ref{pointspec}, we show the spectra of point-like emission from the Galactic Center, as reported in each of these three prior studies. We note that the intensity and spectral features of the Galactic Center point source found by these three groups are very similar, despite the very different analysis techniques employed.

\begin{figure}[t]
\centering
\vspace{0.7cm}
\includegraphics[angle=0.0,width=3.5in]{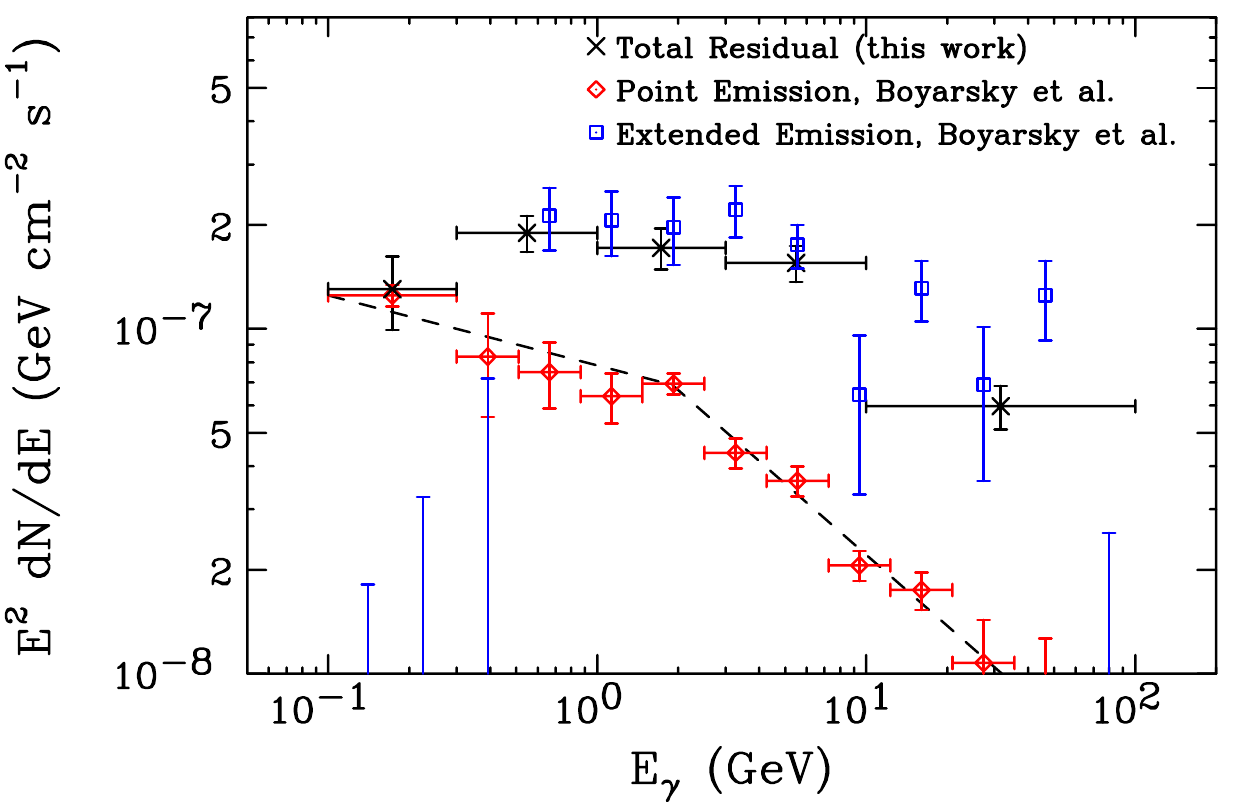}
\caption{A comparison of the total residual emission found in this study (black) with the spectra of point-like emission (red) and extended emission (blue) (as in the case of annihilating dark matter with $\rho_{\rm DM}\propto r^{-1.34}$) as presented in Ref.~\cite{Boyarsky:2010dr}.  This comparison supports our finding that this residual emission below $\sim 300$ MeV is consistent with a point-like source origin, while much of the emission at higher energies is indeed spatially extended.}
\label{compare}
\end{figure}

In Fig.~\ref{compare}, we compare this spectrum of point-like emission (as reported by Boyarsky {\it et al.}~\cite{Boyarsky:2010dr}) to the spectrum of residual emission found in our analysis.\footnote{HESS~\cite{hesspoint} and other ground based telescopes~\cite{others} have also observed point-like emission from the Galactic Center at energies above $\sim$200 GeV. This very high energy gamma ray source may be associated with the point-like emission observed at lower energies, as shown in Fig.~\ref{pointspec}.} Between 100-300 MeV, there is good agreement, indicating that most or even all of the residual gamma rays in this energy range could originate from a single point source. At higher energies, however, the residual emission consistently exceeds the flux attributable to point-like emission; by a factor of $\sim$2-3 between 0.3 and 3 GeV, and by a factor of $\sim$5 above 3 GeV.  When Boyarsky {\it et al.} included a spatially extended component in their model (with a morphology corresponding to that predicted for annihilating dark matter with a distribution given by $\rho_{\rm DM}\propto r^{-1.34}$), they found that the fit improved considerably (reducing the log-likelihood by 25 with the addition of only one new parameter)~\cite{Boyarsky:2010dr}. The spectrum of this spatially extended component is also shown in Fig.~\ref{compare}. The spectrum of the residual emission found in our analysis is in very good agreement with the sum of point-like and extended components as reported by Boyarsky {\it et al}. From these comparisons, we conclude that in addition to the presence of point-like emission from the Galactic Center, a component of extended emission is also prominently present at energies greater than $\sim$300 MeV.\footnote{The spectrum of our residual as presented in Figs.~\ref{spec} and~\ref{compare} denotes the residual within a 5 degree radius around the Galactic Center, whereas the spectrum of extended emission reported in Ref.~\cite{Boyarsky:2010dr} is taken from a similar, but not identical, inner 10$^{\circ}\times 10^{\circ}$ region. Given the highly concentrated nature of the morphology being considered, however, this difference is negligible.}

\section{Possible Origins of the Observed Emission}
\label{origin}

A number of proposals have been put forth to explain the bright gamma ray emission observed from the Galactic Center by the FGST. These possibilities include the central supermassive black hole~\cite{aharonian,HG2}, a population of unresolved millisecond pulsars~\cite{pulsars}, and dark matter annihilations~\cite{HG2,HG1}. In this section, we discuss the morphological and spectral characteristics of the gamma ray emission expected from each of these potential sources and compare this to the emission observed by the FGST.

\subsection{Cosmic ray Acceleration by the Supermassive Black Hole}

To begin, we reiterate that the morphology of the observed emission is not entirely point-like in nature, but instead is somewhat spatially extended. This allows us to rule out the possibility that gamma rays directly emitted by the Milky Way's central black hole are responsible for the observed emission.\footnote{Further supporting this conclusion is the fact that the central emission observed by Fermi shows no variability on month timescales~\cite{aharonian}, as would be expected based on the variability of this source in X-ray and infrared wavelengths~\cite{variability}. The emission from a $\sim$30 parsec source region would have any variability suppressed on timescales shorter than $R/c\sim 100$ years.} If, however, the observed gamma ray spectrum originates from cosmic rays that have been accelerated by the black hole, then a spatially extended distribution of gamma rays could result. 

For example, it was previously proposed that the TeV-scale gamma rays observed from the Galactic Center could originate from the inverse Compton scattering of energetic electrons accelerated by the black hole~\cite{Atoyan:2004ix,Hinton:2006zk}. This scenario, however, predicts considerably less GeV-scale emission than is observed by Fermi, and thus cannot account for the residual emission discussed here~\cite{aharonian}. Alternatively, the black hole may accelerate cosmic ray protons which then diffuse throughout the surrounding interstellar medium, producing pions and thus gamma rays through interactions with gas~\cite{aharonian,aharonianold}. The spectrum and spatial distribution of the gamma ray emission resulting from this process depends not only on the spectrum of protons injected from the black hole, but also on the diffusion coefficient and distribution of gas in the surrounding medium, as well as on the emission history of the black hole (occurrences of flares and periods of relative dormancy). As none of these inputs are currently very well constrained, it is difficult to make reliable predictions for the resulting gamma ray spectrum and distribution. That being said, it appears plausible that a reasonable astrophysical scenario could potentially explain much of the observed emission.

Perhaps the greatest challenge in accounting for the observed emission with energetic protons accelerated by the central black hole is the very rapid increase in the flux of spatially extended emission observed between approximately 200 and 700 MeV (see blue error bars in Fig.~\ref{compare}). Even for a mono-energetic spectrum of protons, the resuling spectrum of gamma rays from pion decay does not rise rapidly enough to account for this feature. Perhaps this could be reconciled, however, if a sizeable fraction of the apparently point-like emission in the 100-300 MeV bin is in fact somewhat extended and arises from cosmic ray interactions.



Lastly, we also note that a sizable fraction of the high energy emission observed by the FGST is likely to be associated with the HESS galactic ridge. This ridge emission, as measured by HESS, possesses a power-law-like spectrum with a spectral index of $2.29\pm0.07_{\rm stat}\pm 0.20_{\rm sys}$ over the energy range of approximately 0.2 to 10 TeV. Due to the spatial correlation of this emission with the locations of molecular clouds in the central 200 parsecs of the Milky Way, the origin of the ridge emission is conventionally taken to be the decays of neutral pions produced in the interactions of cosmic ray protons or nuclei with the surrounding molecular gas. In order to generate a gamma ray spectrum with this spectral index, the responsible protons are required to have a spectral index of approximately $1.9 \pm 0.2$~\cite{HG2}. Based on an extrapolation of this spectral shape, we estimate that on the order of 30\% of the 10-100 GeV residual is associated with the ridge. At energies below $\sim$10 GeV, however, the ridge emission constitutes a much smaller fraction of the observed residual, unless the spectrum of cosmic ray protons in the region is significantly enhanced below $\sim$50 GeV relative to the power-law behavior we have assumed. 





\subsection{Annihilating Dark Matter}

It has long been appreciated that if dark matter particles annihilate in pairs (as predicted in most models of weakly interacting massive particles), the resulting gamma ray signal would be brightest from the direction of the Galactic Center~\cite{gc}. The energy and angular dependent flux of such gamma rays is given by
\begin{equation}
\Phi_{\gamma}(E_{\gamma},\psi) =  \frac{d N_{\gamma}}{d E_{\gamma}} \frac{\sigv}{8\pi m^2_{\rm DM}} \int_{\rm{los}} \rho^2(r) dl,
\label{flux1}
\end{equation}
where $\sigv$ is the dark matter annihilation cross section multiplied by the relative velocity of the two dark matter particles (averaged over the velocity distribution), $m_{\rm DM}$ is the mass of the dark matter particle, $\psi$ is the angle observed relative to the direction of the Galactic Center, $\rho(r)$ is the dark matter density as a function of distance to the Galactic Center, and the integral is performed over the line-of-sight. $d N_{\gamma}/d E_{\gamma}$ is the gamma ray spectrum generated per annihilation, which depends on the mass and dominant annihilation channels of the dark matter particle (we use PYTHIA~\cite{pythia} to calculate $d N_{\gamma}/d E_{\gamma}$ for various dark matter scenarios in this study).

Modern numerical simulations of the evolution of cold dark matter predict the formation of halos with a nearly universal density profile~\cite{Navarro:2008kc}. Within the inner volumes of such halos, the density of dark matter varies as $\rho_{\rm DM} \propto r^{-\gamma}$, where $r$ is the distance to the halo's center. The frequently used Navarro Frenk and White (NFW) profile, for example, features an inner slope of $\gamma=1.0$~\cite{nfw}. The results of the Via Lactea II simulation favor a somewhat steeper inner slope ($\gamma\approx 1.2$)~\cite{vialactea}, while the Aquarius Project finds a somewhat less steep value which varies with $r$~\cite{aquarius}. 

When considering the dark matter distribution in the central kiloparsecs of the Milky Way, it is important to include the effects of stars and gas, which are not taken into account by dark matter-only simulations such as Via Lactea II and Aquarius, but which dominate the gravitational potential of the Inner Galaxy. Generally speaking, as a result of dissipating baryons, dark matter density profiles are expected to be adiabatically contracted, resulting in the steepening of their inner profiles~\cite{ac}. The degree to which this effect is manifest depends on the fraction of the baryons that dissipate slowly by radiative cooling. 

As hydrodynamical simulations which model the process of galaxy formation have improved, efforts to predict the effects of baryons have begun to converge. In particular, several groups (using different codes) have consistently found that Milky Way sized halos are adiabatically contracted, increasing the density of dark matter in their inner volumes relative to that predicted by dark matter-only simulations (see Ref.~\cite{Gnedin:2011uj} and references therein). These simulations, which include the effects of gas cooling, star formation, and stellar feedback, predict a degree of adiabatic contraction which steepens the inner slopes of dark matter density profiles from $\gamma \approx 1.0$ to $\gamma \approx 1.2-1.5$ within the inner $\sim$10 kpc of Milky Way-like galaxies~\cite{Gnedin:2011uj,mac}.  The resolution of such simulations is currently limited to scales larger than $\sim$100 parsecs~\cite{100pc}.

With this information in mind, we can compare the expected spatial distribution of dark matter to the observed angular distribution of gamma rays from around the Galactic Center. Making this comparison, we find that the majority of the residual emission observed between 300 MeV and 10 GeV can be described by annihilating dark matter with a distribution given by $\rho(r)\propto r^{\gamma}$, with $\gamma\approx 1.25-1.40$.\footnote{Below 300 MeV, the observed emission is dominated by point-like emission, and the flux of the emission drops off significantly above 10 GeV, leading us to focus on this energy range.} In contrast, an NFW-like profile with $\gamma=1.0$ would predict a considerably broader distribution of gamma rays than is found in our residual maps. More quantitatively speaking, for $\gamma=1.0$ we find that for energies of 300-1000 MeV, 1-3 GeV, and 3-10 GeV, respectively, no more than 22\%, 18\% and 27\% of the flux found in the innermost half degree around the Galactic Center can arise from dark matter annihilations without also exceeding the flux observed at distances beyond one degree. In contrast, if we select an inner slope of $\gamma=1.3$, we find that up to 72\%, 74\% and 100\% of the innermost emission could originate from dark matter annihilations. The remainder of the residual could easily originate from the central point source with the spectrum presented in Fig.~\ref{pointspec}.

\begin{figure}[t]
\centering
\includegraphics[angle=0.0,width=3.5in]{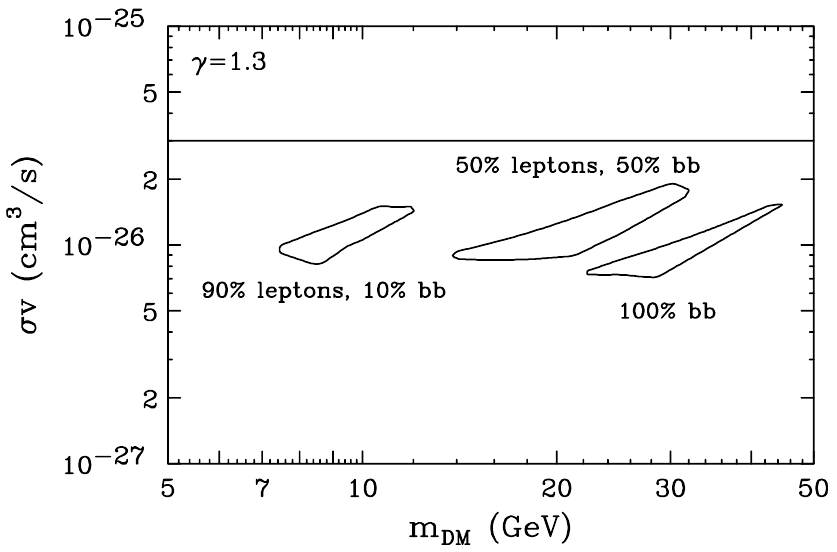}
\caption{The range of dark matter masses and annihilation cross sections for which dark matter annihilations can account for the majority of the observed residual emission between 300 MeV and 10 GeV, for three choices of annihilation channels (``leptons'' denotes equal fractions to $e^+e^-$, $\mu^+\mu^- $ and $\tau^+ \tau^-$). Also shown for comparison is the annihilation cross section predicted for a simple thermal relic ($\sigma v = 3\times 10^{-26}$ cm$^3$/s). Note that there is a factor of a few uncertainty in the annihilation cross section, corresponding to the overall dark matter density and distribution. See text for details.}
\label{fits}
\end{figure}

\begin{figure*}[t]
\centering
\includegraphics[angle=0.0,width=3.5in]{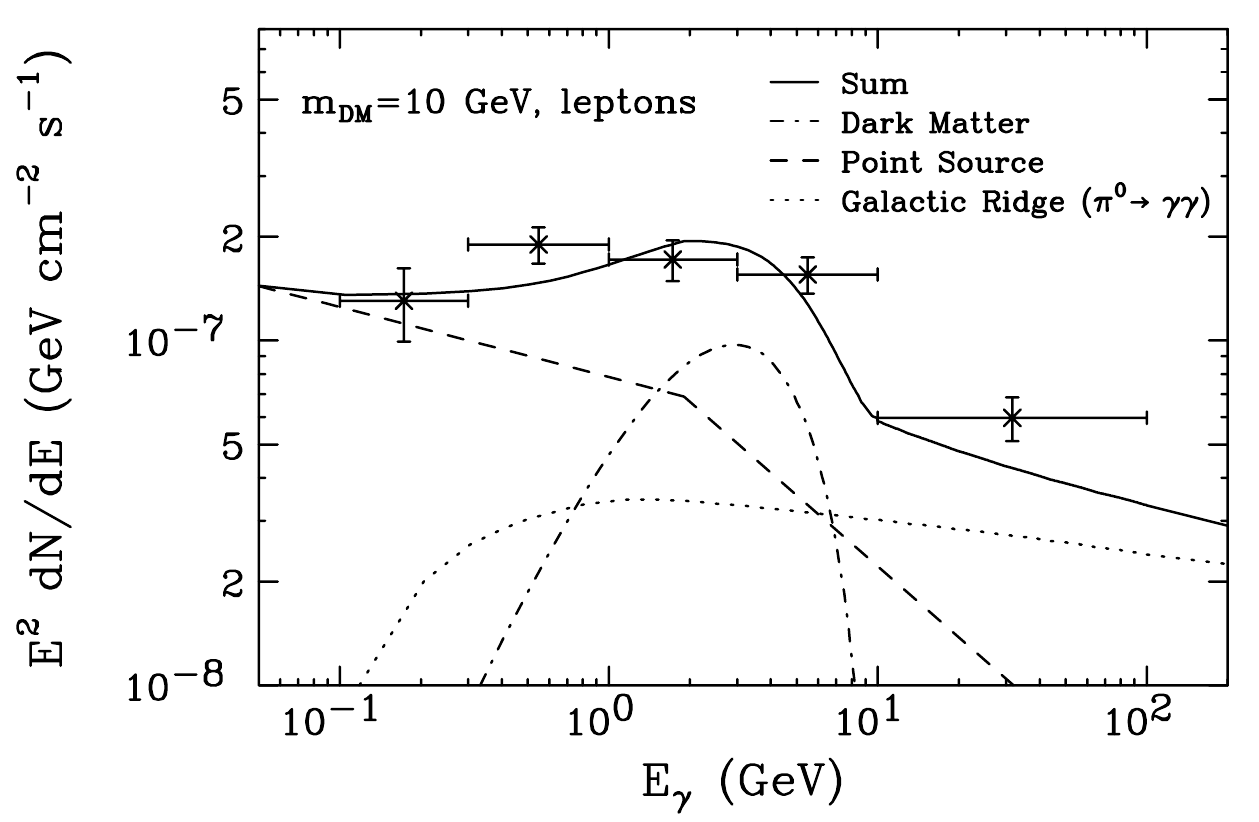}
\includegraphics[angle=0.0,width=3.5in]{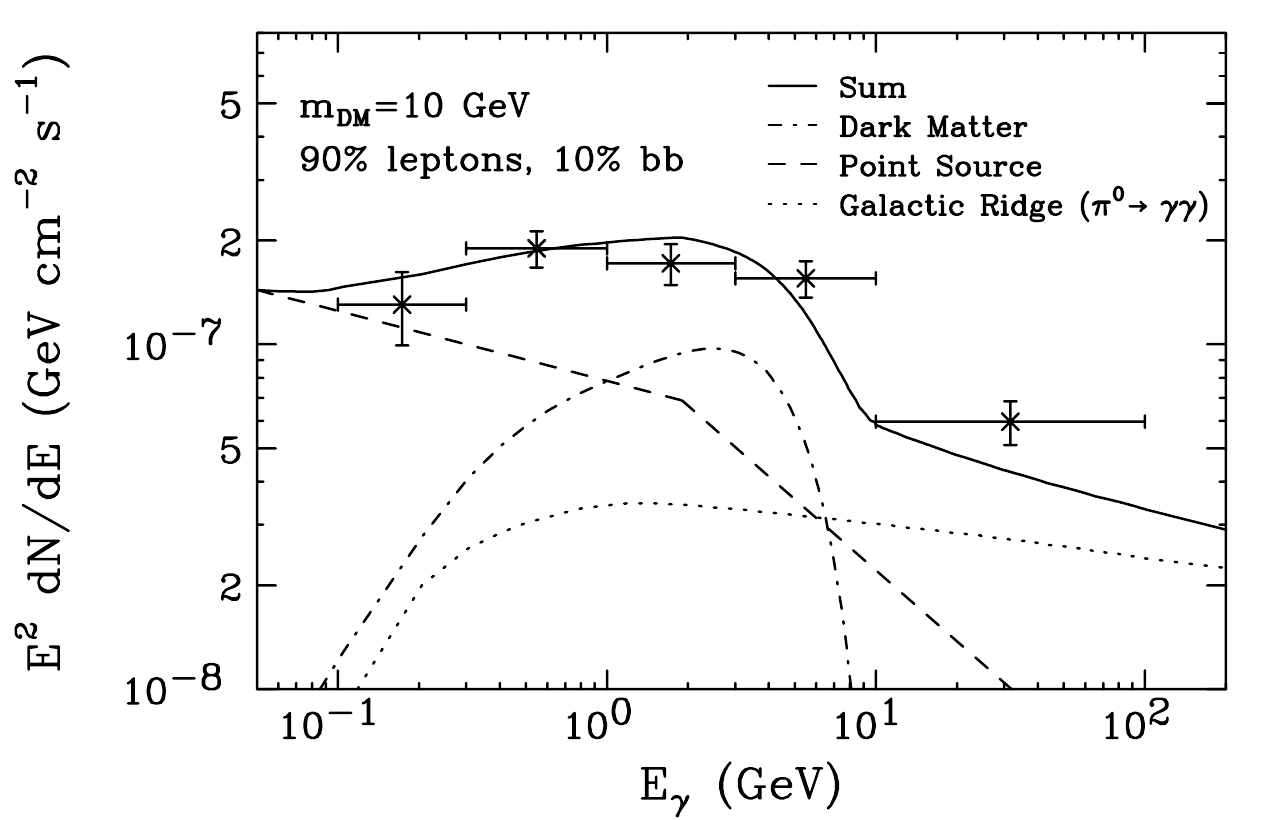}\\
\includegraphics[angle=0.0,width=3.5in]{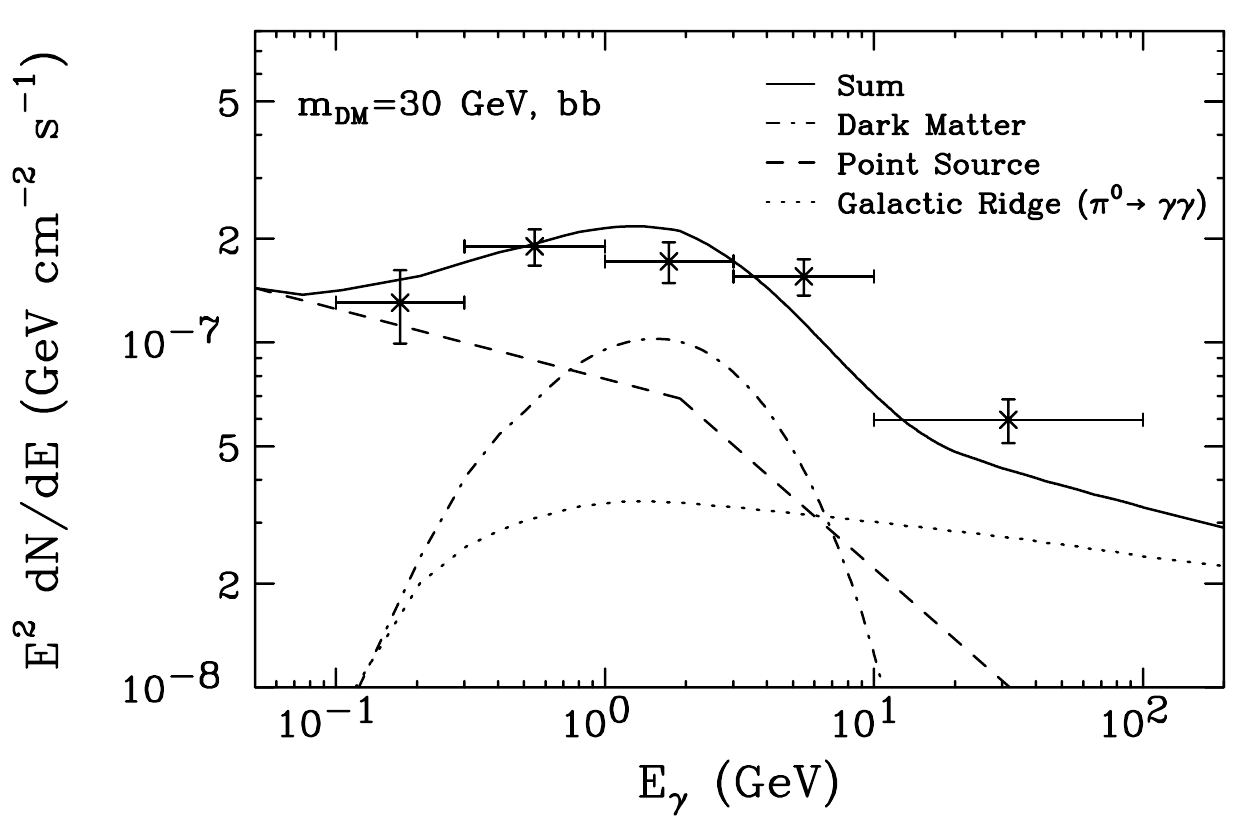}
\caption{Examples illustrating how dark matter annihilations and astrophysical sources could combine to make up the observed residual emission surrounding the Galactic Center. In the upper left frame, we show results for a 10 GeV dark matter particle with an annihilation cross section of $\sigma v = 7\times 10^{-27}$ cm$^3$/s and which annihilates only to leptons ($e^+e^-$, $\mu^+\mu^- $ and $\tau^+ \tau^-$, $1/3$ of the time to each). In the upper right frame, we show the same case, but with 10\% of the annihilations proceeding to $b\bar{b}$. In the lower frame, we show results for a 30 GeV dark matter particle annihilating to $b\bar{b}$ with an annihilation cross section of $\sigma v = 6\times 10^{-27}$ cm$^3$/s. In each case, the annihilation rate is normalized to a halo profile with $\gamma=1.3$. The point source spectrum is taken as the broken power-law shown in Fig.~\ref{pointspec}, and the Galactic Ridge emission has been extrapolated from the higher energy spectrum reported by HESS~\cite{ridge}, assuming a pion decay origin and a power-law proton spectrum. See text for details.}
\label{sum}
\end{figure*}

If a sizable fraction of the residual emission does originate from annihilating dark matter, then we can use the spectrum of this emission to inform us as to the mass and dominant annihilation channels of the dark matter particles. In particular, the rapid decrease in the flux above $\sim$10 GeV suggests that the spectrum is being dominated by $\sim$30 GeV dark matter particles annihilating to quarks, or by $\sim$10 GeV particles annihilating to leptons (among annihilations to leptons, those to taus produce far more gamma rays than those to either muons or electrons). In Fig.~\ref{fits}, we show the range of dark matter masses and annihilation cross sections for which dark matter annihilations can account for the majority of the observed residual emission (without exceeding the observed residual) in each of the three energy bins between 300 MeV and 10 GeV, for three choices of the annihilation channels. Interestingly, we note that the normalization of the signal requires us to consider annihilation cross sections that are within a factor of a few of the value predicted for a simple thermal relic ($\sigma v = 3\times 10^{-26}$ cm$^3$/s). The precise value of the required annihilation cross section depends on the quantity of dark matter present, and is thus subject to the related uncertainties. In Fig.~\ref{fits} and throughout the remainder of this paper, we have normalized the dark matter distribution such that the total mass of dark matter within the solar circle is $3.76\times 10^{67}$ GeV, which is the value corresponding to the case of $\gamma=1.0$ and a local density of 0.4 GeV/cm$^3$. This value is supported by a combination of microlensing and dynamical constraints, although uncertainties exists~\cite{local}. With these uncertainties in mind, one should consider all annihilation cross sections shown in Fig.~\ref{fits} and elsewhere in this paper to be accurate only to within a factor of a few.

Of course, it is also expected that astrophysical sources will contribute to the Galactic Center's gamma ray spectrum between 300 MeV and 10 GeV. In Fig.~\ref{sum}, we show three examples in which emission from a central point source (as shown in Fig.~\ref{pointspec}), along with emission from the Galactic Ridge (as extrapolated from the higher energy HESS emission, assuming a spectral shape that results from a power-law spectrum of protons) combine with a contribution from dark matter to generate the observed residual emission. Note that the lowest energy emission is largely generated by the central point source (as suggested by the observed morphology) while the highest energy bin is dominated by emission from the Galactic Ridge. Only the 300 MeV-10 GeV range is dominated by dark matter annihilation products.

\subsection{Millisecond Pulsars}

A population of gamma ray point sources surrounding the Galactic Center could also potentially contribute to the observed residual emission. Millisecond pulsars, which are observed to produce spectra that fall off rapidly above a few GeV, represent such a possibility~\cite{HG2,pulsars}.

\begin{figure}[t]
\centering
\vspace{0.3in}
\includegraphics[angle=0.0,width=3.5in]{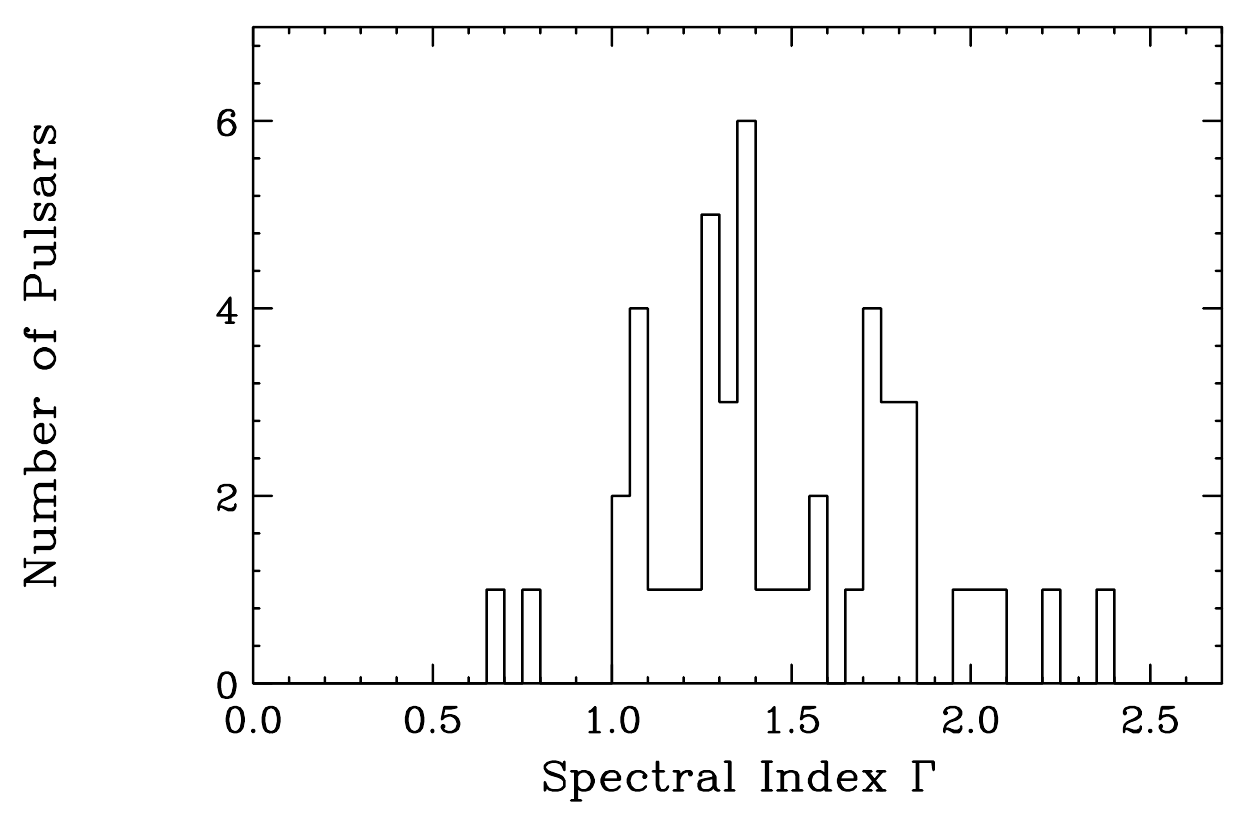}
\caption{A histogram showing the distribution of spectral indices, $\Gamma$, of pulsars in the Fermi Pulsar Catalog.}
\label{pulsar}
\end{figure}

Observations of resolved millisecond pulsars by FGST have found an average spectrum well described by $dN_\gamma/dE_{\gamma} \propto E_{\gamma}^{-1.5} \exp(-E_{\gamma}/2.8\,{\rm GeV})$~\cite{averagepulsar}. Similarly, the 46 gamma ray pulsars (millisecond and otherwise) in the FGST's first pulsar catalog have a distribution of spectral indices which peaks strongly at $\Gamma=$1.38, with 44 out of 46 of the observed pulsars possessing (central values of their) spectral indices greater than 1.0~\cite{pulsarcatalog} (see Fig.~\ref{pulsar}). In contrast, to produce a sizable fraction of the spatially extended residual emission between 300 MeV and 10 GeV without exceeding the emission observed below 300 MeV, the average pulsar in the Galactic Center population would be required to possess a spectral index harder than $\Gamma\approx 1.0$. And although we agree with the author of Ref.~\cite{pulsars} that a small number of pulsars (including J1958+2846, J2032+4127 and J2043+2740) have been observed with such hard spectral indices, we do not believe that the existing data supports the conclusion that a large population of pulsars (as would be required to generate the observed emission) would produce an average gamma ray flux with a spectral shape able to account for the observed emission from the Galactic Center.\footnote{The error bars on the spectral indices of these three hardest pulsars are also quite large, $\Gamma=0.77\pm0.31$, $0.68\pm0.46$, and $1.07\pm0.66$~\cite{pulsarcatalog}.} That being said, if the population of pulsars present in the central stellar cluster were to differ significantly from the sample represented by the Fermi pulsar catalog, a different conclusion could potentially be reached.

\begin{figure}[t]
\centering
\includegraphics[angle=0.0,width=3.3in]{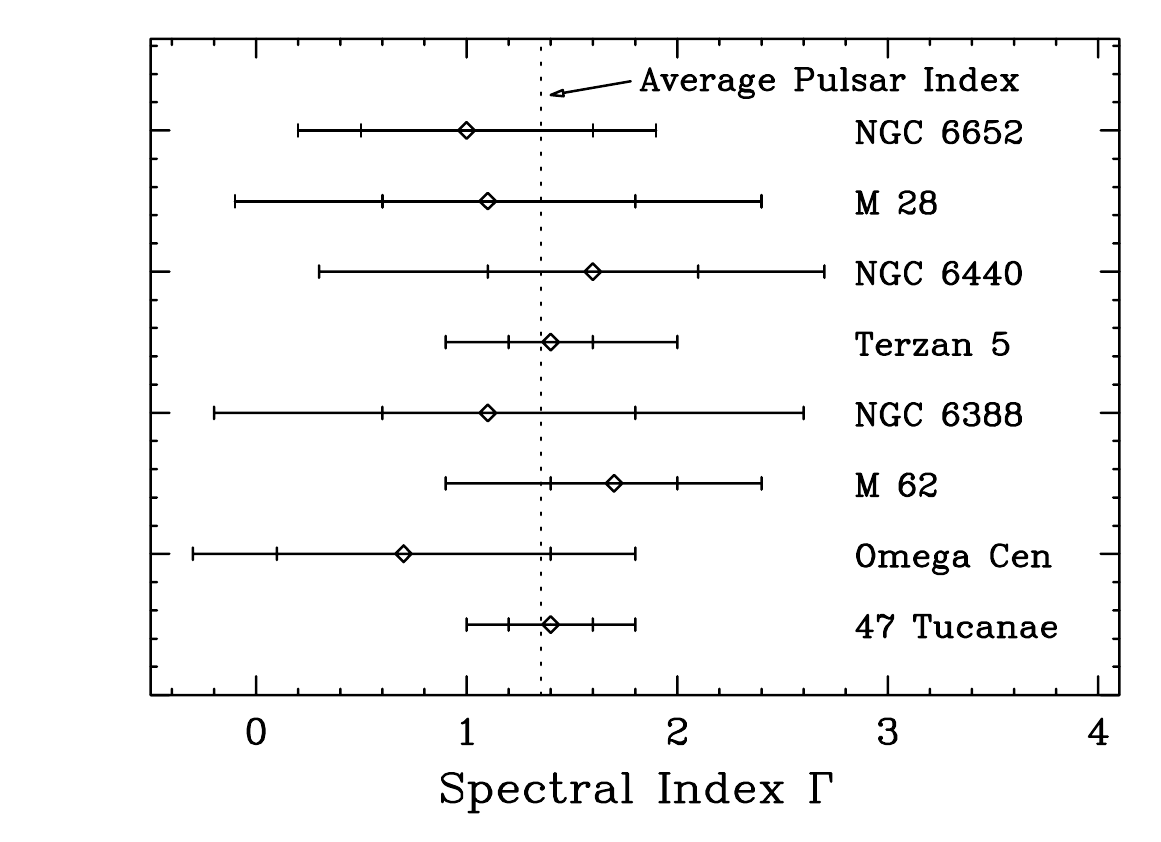}
\caption{The spectral indices (with statistical and systematic error bars) of the eight globular clusters observed by the Fermi Gamma Ray Space Telescope~\cite{glob}.}
\label{globular}
\end{figure}

An opportunity to measure the emission from large populations of gamma ray pulsars exists in the form of globular clusters, whose gamma ray emission is generally attributed to pulsars contained within their volumes. Unfortunately, the gamma ray spectra of these objects have not been well measured. In particular, the eight globular clusters with spectra reported by Fermi have an average spectral index very close to that of pulsars ($\Gamma \approx 1.38$), but with very large individual error bars which extend from roughly 0 to 2.5 (these values, including 1$\sigma$ statistical and systematic errors are shown in Fig.~\ref{globular}). Perhaps with more data, we will learn from these systems whether the spectral indices of large pulsar populations can be hard enough to accommodate the emission observed from the Galactic Center.

Lastly, we note that it is somewhat difficult to accommodate the very spatially concentrated morphology of the observed gamma ray emission with pulsars. As originally pointed out in Ref.~\cite{HG2}, to match the observed angular distribution of this signal, the number density of pulsars would have to fall off with the distance to the Galactic Center at least as rapidly as $r^{-2.5}$. In contrast, within the innermost parsec of the Galactic Center, the stellar density has been observed to fall off only about half as rapidly, $r^{-1.25}$~\cite{Schoedel:2009mv}. Furthermore, even modest pulsar kicks of $\sim 100$ km/s would allow a pulsar 10 pc from the Galactic Center to escape the region, consequently broadening the angular width of the signal. Annihilating dark matter, in contrast, produces a flux of gamma rays that scales with its density {\it squared}, and thus can much more easily account for the high concentration of the observed signal.

\section{Constraints On The Dark Matter Annihilation Cross Section}
\label{con}

In this section, instead of attempting to determine the origin of the gamma rays from the Galactic Center region, we use the observed spectrum and flux to place limits on the dark matter annihilation cross section. In doing this, we do not assume anything about the source or sources responsible for the observed emission, but instead only require that dark matter annihilation products do not exceed the observed emission (after subtracting the known sources and line-of-sight gas templates, as described in Sec.~\ref{analysis}). Despite using this very simple and conservative approach, we derive constraints that are competitive with or stronger than those placed by other indirect search strategies, including those from observations of dwarf spheroidals~\cite{dwarf}, galaxy clusters~\cite{clusters}, the cosmological diffuse background~\cite{cosmo}, and nearby subhalos~\cite{subhalos}.

\begin{figure*}[t]
\centering
\includegraphics[angle=0.0,width=5.4in]{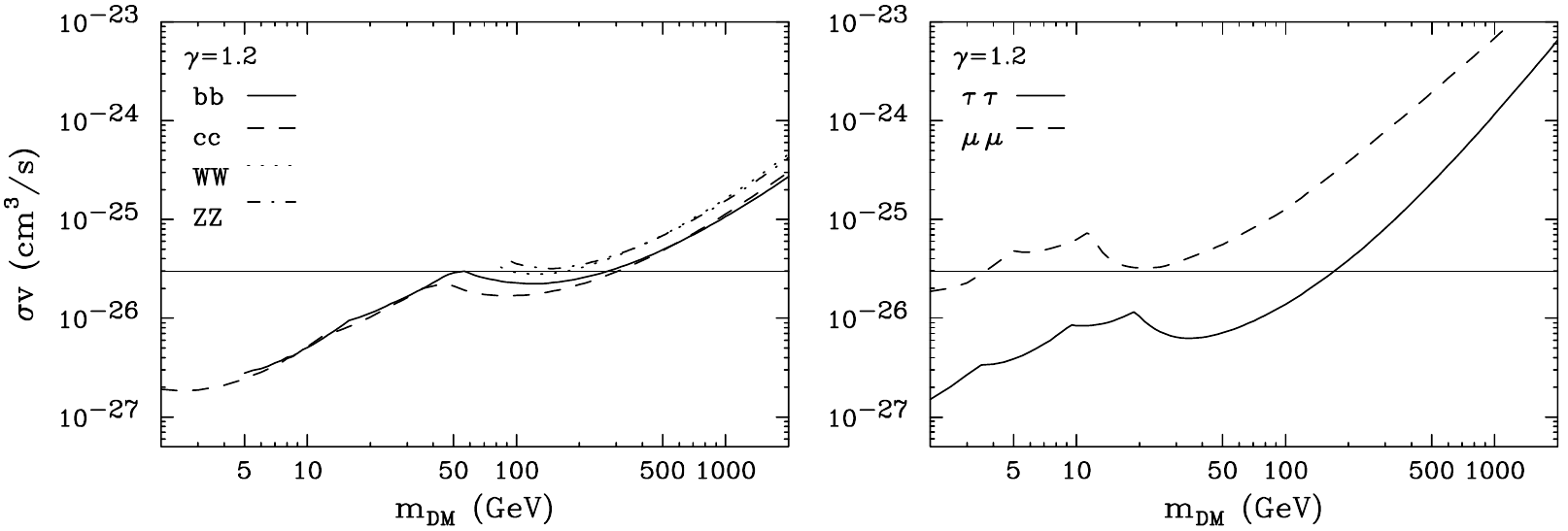}\\
\includegraphics[angle=0.0,width=5.4in]{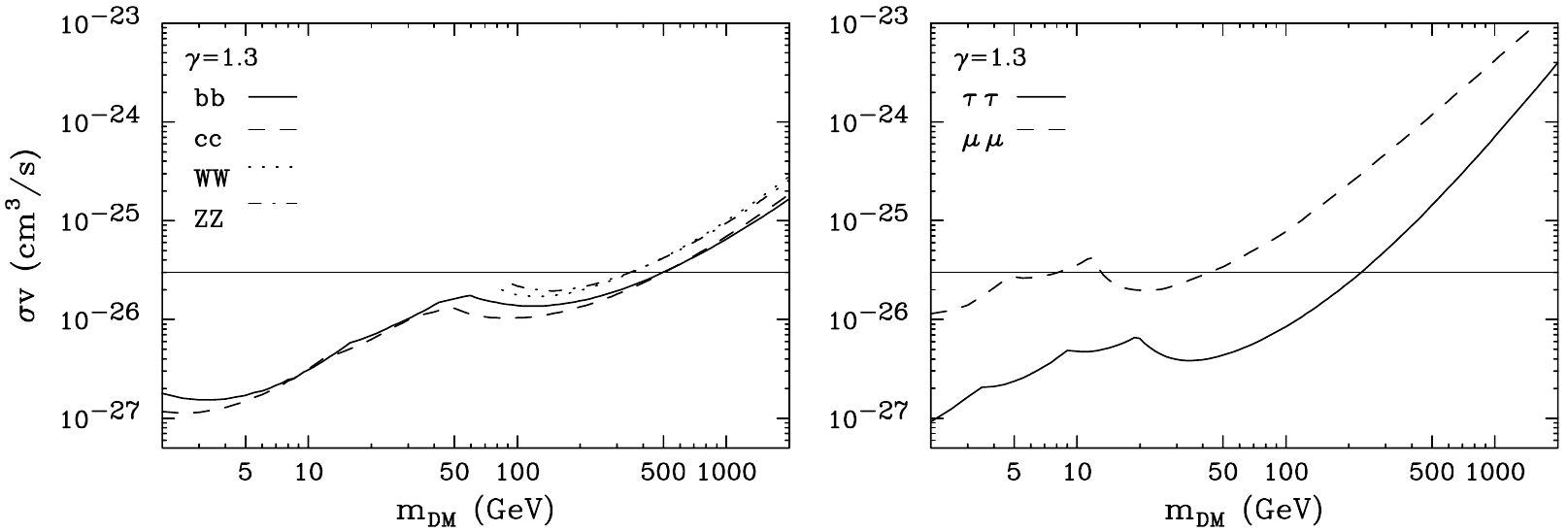}\\
\includegraphics[angle=0.0,width=5.4in]{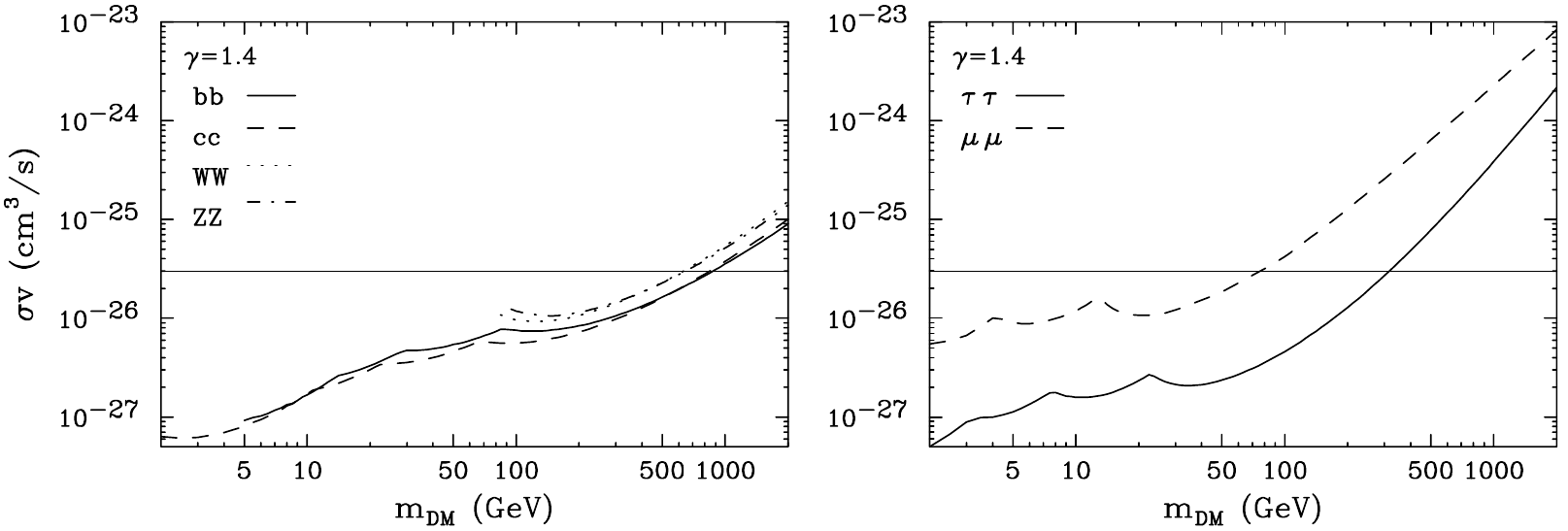}
\caption{Upper limits on the dark matter annihilation cross section for several choices of the final state and for three values of the halo profile's inner slope, $\gamma$. Also shown for comparison is the annihilation cross section predicted for a simple thermal relic ($\sigma v = 3\times 10^{-26}$ cm$^3$/s). Uncertainties in the overall dark matter density have not been included, but based on the errors presented in Ref.~\cite{local}, we expect that this would only weaken our limits by about 30-50\%. See text for more details.}
\label{constraints}
\end{figure*}

In Fig.~\ref{constraints}, we show the 95\% confidence level upper limits on the dark matter annihilation cross section for several choice of the final state, and for three values of the halo profile's inner slope. Based on the results of hydrodynamical simulations~\cite{Gnedin:2011uj,mac}, we consider the $\gamma=1.2$ to represent the minimal degree of baryonic contraction, where as the $\gamma=1.3$ and 1.4 cases should be taken as more central estimates. 

The constraints we have derived from the Galactic Center region are indeed quite stringent. Even in the case of only a very modest degree of baryonic contraction ($\gamma=1.2$), we find that dark matter particles with the canonical annihilation cross section of $\sigma v = 3\times 10^{-26}$ cm$^3$/s and which proceed to hadronic final states are predicted to exceed the observed gamma ray flux from the Galactic Center unless they are more massive than approximately 300 GeV. In comparison, the Fermi collaboration's combined analysis of 10 dwarf spheroidals only excludes such dark matter particles with masses below approximately 30 GeV~\cite{dwarf}.\footnote{Unlike the central regions of Milky Way-like halos, the dark matter density profiles of dwarf spheroidal galaxies are not generally expected to be contracted by baryons~\cite{dwarfcores}. The uncontracted NFW profile adopted in the Fermi Collaboration dwarf spheroidal analysis is thus appropriate.}

\begin{figure*}[t]
\centering
\includegraphics[angle=0.0,width=5.4in]{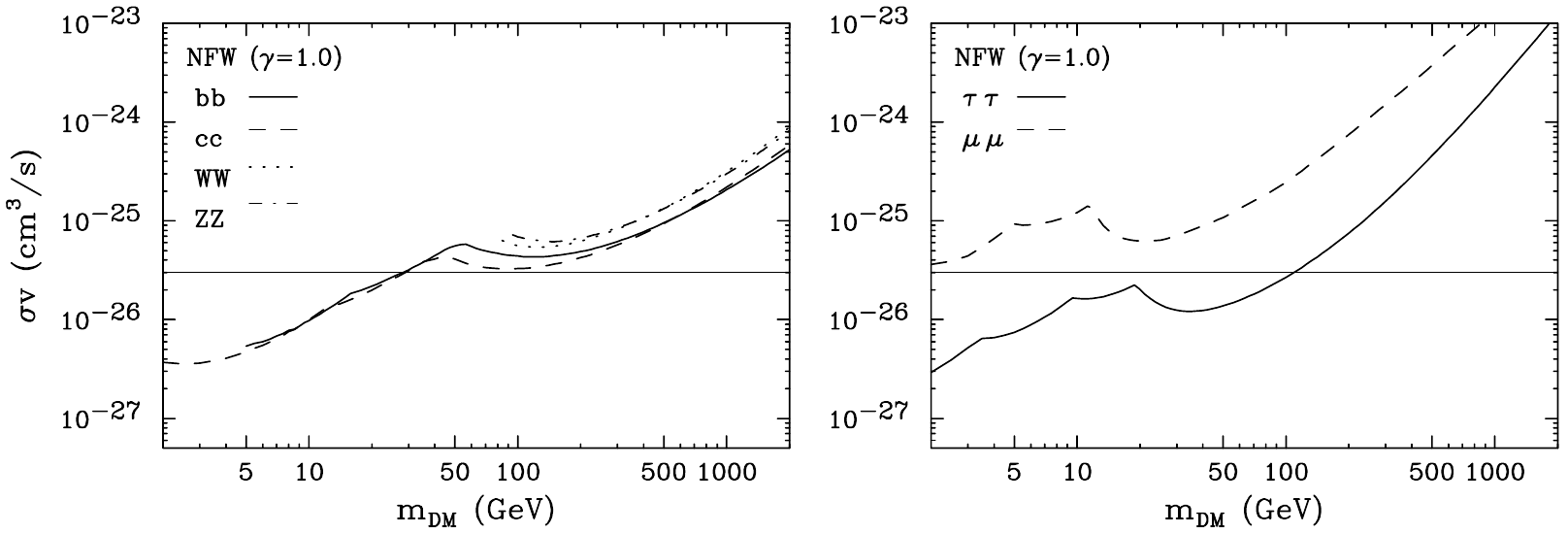}
\caption{Upper limits on the dark matter annihilation cross section for several choice of the final state, neglecting the effects of baryons (using an uncontracted NFW halo profile). Also shown for comparison is the annihilation cross section predicted for a simple thermal relic ($\sigma v = 3\times 10^{-26}$ cm$^3$/s). Uncertainties in the overall dark matter density have not been included, but based on the errors presented in Ref.~\cite{local}, we expect that this would only weaken our limits by about 30-50\%. See text for more details.}
\label{constraintsnfw}
\end{figure*}

In Fig.~\ref{constraintsnfw}, we also show the constraints which result if the effects of baryons on the dark matter distribution are neglected (using a uncontracted NFW profile). Even in this case, we find limits which are approximately as stringent as those derived from dwarf galaxies.

\section{Discussion and Summary}
\label{summary}

In this article, we have used the first three years of data taken by the Fermi Gamma Ray Space Telescope (FGST) to study the spectrum and spatial morphology of the gamma ray emission from the region surrounding the Galactic Center. In doing so, we have identified a spatially extended component of gamma ray emission which peaks at energies between approximately 300 MeV and 10 GeV. The origin of these gamma rays is currently uncertain, although they could potentially be the annihilation products of dark matter particles, or the products of collisions of high energy protons accelerated by the Milky Way's supermassive black hole with gas. 

If this extended source of gamma rays is interpreted as dark matter annihilation products, the spectrum of this emission favors dark matter particles with a mass in the range of 7-12 GeV (if annihilating dominantly to leptons) or 25-45 GeV (if annihilating dominantly to hadronic final states). The former of these mass ranges is of particular interest in light of the observations reported by the direct detection experiments DAMA/LIBRA~\cite{dama}, CoGeNT~\cite{cogent}, and CRESST~\cite{cresst}, which each report signals consistent with an approximately 10 GeV dark matter particle (see also, however, constraints from the CDMS~\cite{cdms} and XENON~\cite{xenon} collaborations, and related discussions~\cite{juan}). Further motivating the dark matter interpretation of the Galactic Center gamma rays is the fact that the annihilation cross section required to normalize the annihilation rate to the observed flux is approximately equal to the value required to generate the observed cosmological abundance in the early universe ($\sigma v\sim 3 \times 10^{-26}$ cm$^3$/s). In other words, in lieu of resonances, coannihilations, P-wave suppression, or other complicating factors, a particle species that will freeze-out in the early universe with a density equal to the measured dark matter abundance is also predicted to annihilate today at a rate that is similar to that needed to produce the observed gamma rays from the Galactic Center. 

Additionally, we point out that if dark matter particles are annihilating in the Inner Galaxy at the rate required to produce the observed gamma ray flux, then the resulting energetic electrons and positrons will diffuse outward, potentially producing observable quantities of synchrotron emission. In particular, focusing on the case of 7-12 GeV dark matter particles annihilating dominantly to leptons, the halo profile and cross section required to produce the morphology and normalization of the observed gamma ray flux is also predicted to lead to the production of a diffuse haze of synchrotron emission very similar to that observed by WMAP~\cite{haze} (see Fig.~3 of Ref.~\cite{timhaze} for a direct comparison). It also appears that the excess radio emission observed at higher galactic longitudes by the ARCADE 2 experiment~\cite{arcade} possesses a spectral shape and overall intensity consistent with originating from dark matter with the same mass, cross section, dominant channels, and distribution~\cite{arcadedarkmatter,pc}. Lastly, we mention that 7-12 GeV dark matter particles with the distribution and annihilation cross section favored here would be capable of depositing the required energetic electrons into the Milky Way's non-thermal radio filaments~\cite{filaments}, providing an explanation for their peculiar spectral features.

It is noteworthy that the different explanations proposed for the observed gamma ray emission from the Galactic Center predict different accompanying spectra of cosmic ray electrons, potentially providing us with a way to discriminate between these different scenarios. Of the sources proposed for the observed gamma ray emission, only dark matter annihilations are predicted to produce comparable fluxes of gamma rays and electrons, with spectra that peak at similar energies. Pulsars, in contrast, produce gamma ray spectra which peak at $\sim$1-3 GeV and electron spectra which peak at several hundred GeV~\cite{pulsarelectrons}. Perhaps future observations of the Inner Galaxy at radio and microwave frequencies will be able to make use of this comparison to shed light on the origin of the gamma ray emission from the center of our galaxy.

Lastly, we have also presented conservative limits on the dark matter's annihilation cross section which are at least as stringent as those derived from other observations, such as those of dwarf spheroidal galaxies.

\bigskip
\bigskip

{\it Acknowledgements}: We would like to thank Kev Abazajian, Alyson Brooks, Greg Dobler, Scott Dodelson, Will Farr, Doug Finkbeiner, Nick Gnedin, Jeter Hall, Jim Hinton, Stefano Profumo, and Marco Regis for helpful comments and discussions. DH is supported by the US Department of Energy and by NASA grant NAG5-10842.

\end{document}